\documentclass[fleqn,usenatbib]{mnras}

\usepackage{newtxtext,newtxmath}

\usepackage[T1]{fontenc}

\DeclareRobustCommand{\VAN}[3]{#2}
\let\VANthebibliography\thebibliography
\def\thebibliography{\DeclareRobustCommand{\VAN}[3]{##3}\VANthebibliography}

\usepackage{graphicx}	
\usepackage{amsmath}	

\newcommand{\nustar}{{\it NuSTAR}}
\newcommand{\swift}{{\it Swift}}
\newcommand{\xmm}{{\it XMM-Newton}}
\newcommand{\eps}{erg s$^{-1}$}
\newcommand{\ecs}{erg cm$^{-2}$ s$^{-1}$}
\newcommand{\pcm}{cm$^{-2}$}
\newcommand{\source}{NGC~1566}
\newcommand{\M}{$M_{\odot}$}
\newcommand{\phc}{ph cm$^{-2}$ s$^{-1}$}

\title[The Changing-Look AGN NGC~1566]{Broadband X-Ray Observations of the 2018 Outburst of the Changing-Look Active Galactic Nucleus NGC~1566}

\author[A. Jana et al.]{
Arghajit Jana,$^{1}$\thanks{E-mail: argha@prl.res.in}
Neeraj Kumari,$^{1,2}$
Prantik Nandi,$^{3}$
Sachindra Naik,$^{1}$
Arka Chatterjee,$^{4,5}$
Gaurava K. Jaisawal,$^6$
\newauthor
Kimitake Hayasaki,$^5$ 
Claudio Ricci$^{7,8}$
\\
$^{1}$Astronomy \& Astrophysics Division, Physical Research Laboratory, Navrangpura, Ahmedabad, 380009, India\\
$^{2}$Department of Physics, Indian Institute of Technology, Gandhinagar- 382355, Gujarat, India\\
$^{3}$Department of Astrophysics and Cosmology, S. N. Bose National Centre for Basic Science, Block-JD, Sector-III, Salt Lake, Kolkata, 700106, India\\
$^{4}$Department of Physics, School of Natural Sciences, UNIST, Ulsan, 44919, Republic of Korea\\
$^{5}$Department of Astronomy and Space Science, Chungbuk National University, Cheongju, 361-763, Republic of Korea\\
$^6$National Space Institute, Technical University of Denmark, Elektrovej 327-328, DK-2800 Lyngby, Denmark\\
$^7$N{\'u}cleo de Astronom{\'i}a de la Facultad de Ingenier{\'i}a, Universidad Diego Portales, Av. Ej{\'e}rcito Libertador 441, Santiago, Chile\\
$^{8}$Kavli Institute for Astronomy and Astrophysics, Peking University, Beijing 100871, People's Republic of China\\
}
\date{Accepted XXX. Received YYY; in original form ZZZ}

\pubyear{2021}

\begin{document}
\label{firstpage}
\pagerange{\pageref{firstpage}--\pageref{lastpage}}
\maketitle

\begin{abstract}
We study the nature of the changing-look Active Galactic Nucleus NGC~1566 during its June 2018 outburst. During the outburst, the X-ray intensity of the source rises up to $\sim 25-30$ times compared to its quiescent state intensity. We perform timing and spectral analysis of the source during pre-outburst, outburst and post-outburst epochs using semi-simultaneous observations with the {\it XMM-Newton}, {\it Nuclear Spectroscopic Telescope Array (\nustar)} and {\it Neil Gehrels Swift} Observatories. We calculate variance, normalized variance, and fractional rms amplitude in different energy bands to study the variability. The broad-band 0.5--70 keV spectra are fitted with phenomenological models, as well as physical models. A strong soft X-ray excess is detected in the spectra during the outburst. The soft excess emission is found to be complex and could originate in the warm Comptonizing region in the inner accretion disc. We find that the increase in the accretion rate is responsible for the sudden rise in luminosity. This is supported by the `q'-shape of the hardness-intensity diagram that is generally found in outbursting black hole X-ray binaries. From our analysis, we find that NGC~1566 most likely harbours a low-spinning black hole with the spin parameter  $a^{*} \sim 0.2$. We also discuss a scenario where the central core of NGC~1566 could be a merging supermassive black hole.
\end{abstract}



\label{sec:intro}
\section{Introduction}
Active galactic nuclei (AGNs) are classified as type-1 or type-2, depending on the presence or absence of broad optical emission lines. The existence of different classes of AGNs can be explained by the unified model \citep{Antonucci1993}, which is based on the orientation of the optically thick torus with respect to our line-of-sight. Recently, a new sub-class of AGNs, known as changing look AGNs (CLAGNs), has been identified by optical observations. These objects display the appearence or disappearance of the broad optical emission lines, transitioning from type-1 (or type~1.2/1.5) to type-2 (or type~1.8/1.9) and vice versa. Several nearby galaxies, such as Mrk~590 \citep{Denney2014}, NGC~2617 \citep{Shappee2014}, Mrk~1018 \citep{Cohen1986}, NGC~7582 \citep{Aretxaga1999}, NGC~3065 \citep{Eracleous2001}, have been found to show such a peculiar behaviour. In the X-rays,  a different type of changing-look events have been observed, with AGN switching between Compton-thin (line-of-sight column density, $N_{\rm H} < 1.5 \times 10^{24}$ \pcm) and Compton-thick (CT; $N_{\rm H} > 1.5 \times 10^{24}$ \pcm) states \citep{Risaliti2002,Matt2003}. These X-ray changing-look events have been observed in many AGNs, namely, NGC~1365 \citep{Risaliti2007}, NGC~4388 \citep{Elvis2004}, NGC~7582 \citep{Piconcelli2007,Bianchi2009}, NGC~4395 \citep{Nardini2011}, IC~751 \citep{Ricci2016}, NGC~4507 \citep{Braito2013}, NGC~6300 \citep{Guainazzi2002,AJ2020}.

The origin of the CL events is still unclear. The X-ray changing-look events could be explained by variability of the line-of-sight column density ($N_{\rm H}$) associated with the clumpiness of the BLR or of the circumnuclear molecular torus \citep{Nenkova2008a,Nenkova2008b,Elitzur2012,Yaqoob2015,Guainazzi2016,AJ2020}. On the other hand, optical CL events could be related to changes in the accretion rate \citep{Elitzur2014,MacLeod2016,Sheng2017}, which could be linked with the appearance and disappearance of the broad line regions (BLRs) \citep{Nicastro2000,Korista2004,Runnoe2016}. Some of these optical CL events could also be associated with the tidal disruption of a star by the supermassive black hole (SMBH) at the centre of the galaxy \citep{Eracleous1995,Merloni2015,Ricci2020,Ricci2021}.

NGC~1566 is a nearby (z=0.005), face-on spiral galaxy, classified as a type~SAB(s)bc \citep{devaucouleurs1973,Shobbrook1966}. The AGN was intensively studied over the last 70 years and is one of the first galaxies where variability was detected \citep{Quintana1975,Alloin1985,Alloin1986,Winkler1992}. In the 1960s, \source~ was classified as a Seyfert~1, with broad H$\alpha$ and H$\beta$ lines \citep{devaucouleurs1961}. Later, the H$\beta$ line was found to be weak, leading to the source being classified as a Seyfert~2 \citep{Pastoriza1970}. In the 1970s and 1980s, \source~ was observed to be in the low state with weak H$\beta$ emission \citep{Alloin1986}. Over the years, it was observed to change its type again from Seyfert~1.9-1.8 to Seyfert~1.5-1.2 \citep{dasilva2017}, with two optical outbursts in 1962 and 1992 \citep{Shobbrook1966,Pastoriza1970,dasilva2017,Oknyansky2019}.

{\it INTEGRAL} caught \source~ in the outburst state in hard X-ray band in June 2018 \citep{Ducci18}. Follow-up observations were carried out in the X-ray, optical, ultraviolet (UV), and infrared (IR) bands  \citep{Grupe18a,Ferrigno18,Kuin18,Dai18,Cutri18}. The flux of the AGN was found to increase in all wavebands and reached its peak in July 2018 \citep{Oknyansky2019,Oknyansky2020,Parker2019}. Long-term ASAS-SN and NEOWISE light curves showed that the optical and IR flux started to increase from September 2017 \citep{Dai18,Cutri18}. The {\it Swift}/XRT flux increased by about $\sim 25-30$ times \citep{Oknyansky2019} as the source changed to Seyfert~1.2 from Seyfert~1.8-1.9 type \citep{Oknyansky2019,Oknyansky2020}. The source became a type-1, with the appearance of strong, broad emission lines \citep{Oknyansky2019,Ochmann2020}. After reaching their peak, the fluxes declined in all wavebands. After the main outburst, several small flares were observed \citep{Grupe18b,Grupe19}.

In this paper, we explore the timing and spectral properties of NGC~1566 during the 2018 outburst using data from the \xmm~ and \nustar~ observatories, covering a broad energy range ($0.5-70$~keV). In Section~2, we present the observations and describe the procedure adopted for the data extraction. In Section~3 \& 4, we present results obtained from our timing and spectral analysis, respectively. In Section~5, we discuss our findings. We summarise our results in Section~6.

\begin{table*}
\caption{Log of observations of NGC~1566}
\label{tab:log}
\begin{tabular}{lcccccccccc}
\hline
ID & UT Date & \nustar~ & Exp (ks) & Count s$^{-1}$  & \xmm~ & Exp (ks) & Count s$^{-1}$ & \swift/XRT & Exp (ks) & Count s$^{-1}$ \\
\hline
X1 & 2015-11-05 & --          & --    & --  & 0763500201 & 91.0 & $1.21\pm0.01$  & --          &  --  &     -- \\
O1 & 2018-06-26 & 80301601002 & 56.8 & $1.99\pm0.01$ & 0800840201 & 94.2 & $26.40\pm0.02$  & --          &  --  &     -- \\
O2 & 2018-10-04 & 80401601002 & 75.4 & $0.55\pm0.003$ & 0820530401 & 108.0& $4.92\pm0.01$  & --          &  --  &     -- \\
O3 & 2019-06-05 & 80502606002 & 57.2 & $0.19\pm0.002$ & 0840800401 & 94.0 & $3.78\pm0.01$  & --          &  --  &     -- \\
O4 & 2019-08-08 & 60501031002 & 58.9 & $0.60\pm0.003$ & --         & --    & --     & 00088910001 & 1.9 & $0.55\pm0.02$ \\
O5 & 2019-08-11 & --          & --    & --     & 0851980101 & 18.0 & $4.83\pm0.02$ & --          &  --  &     -- \\
O6 & 2019-08-18 & 60501031004 & 77.2 &$0.33\pm0.002$ & --         & --    & --     & 00088910002 & 1.7 & $0.27\pm0.01$ \\
O7 & 2019-08-21 & 60501031006 & 86.0 & $0.36\pm0.002$ & --         & --    & --     & 00088910003 & 2.0 & $0.21\pm0.01$ \\
\hline
\end{tabular}
\end{table*}

\label{sec:obs}
\section{Observation and Data Reduction}
In the present work, we used data from \nustar, \xmm~ and \swift~ observations of NGC~1566, carried out at different epochs, as reported in Table~\ref{tab:log}. Out of the eight available epochs, X1 and O1 were studied by \citet{Parker2019}. We also included those observations for a complete study of the source at different phases (pre-outbursts, outburst and post-outbursts period).

\label{sec:nustar}
\subsection{\nustar}

\nustar~ is a hard X-ray focusing telescopes, consisting of two identical modules: FPMA and FPMB \citep{Harrison2013}. \source~ was observed by \nustar~ six times between 2018 June 26 and 2019 August 21, simultaneously with either \xmm~ or \swift~ (see Table~\ref{tab:log}). Reprocessing of the raw data was performed with the \nustar~ Data Analysis Software ({\tt NuSTARDAS}, version 1.4.1). Cleaned event files were generated and calibrated by using the standard filtering criteria in the {\tt nupipeline} task and the latest calibration data files available in the NuSTAR calibration database (CALDB) \footnote{\url{http://heasarc.gsfc.nasa.gov/FTP/caldb/data/nustar/fpm/}}. The source and background products were extracted by considering circular regions with radius 60 arcsec and 90 arcsec, respectively. While the source region was centred at the coordinates of the optical counterparts, the background spectrum was extracted from a region devoid of other sources. The spectra and light curves were extracted using the {\tt nuproduct} task. The light curves were binned at 500~s. We re-binned the spectra with 20 counts per bin by using the {\tt grppha} task. No background flare was detected in the \nustar~ observations.

\label{sec:xmm}
\subsection{\xmm}

NGC~1566 was observed by \xmm~ \citep{Jansen2001} at five epochs between November 2015 and August 2019. Out of these five observations, the source was observed simultaneously with \nustar~ in three epochs (see Table~\ref{tab:lag}). We used the Science Analysis System ({\tt SAS v16.1.0}\footnote{\url{https://www.cosmos.esa.int/web/xmm-newton/sas-threads}}) to reprocess the raw data from EPIC-pn \citep{Struder2001}. We considered only unflagged events with {\tt PATTERN} $\leq 4$. Particle background flares were observed above 10 keV in X1 and O5. The Good Time Interval (GTI) file was generated considering only intervals with $<0.65$ counts$^{-1}$, using the SAS task {\tt tabgtigen}. No flares were observed in O1, O2 and O3. The source and background spectra were initially extracted from a circular region of 30" centered at the position of the optical counterpart, and from a circular region of 60" radius away from the source, respectively. Then we extracted the spectrum using the SAS task `{\tt especget}'. The observed and expected pattern distributions were generated by applying {\tt epaplot} task on the filtered events. We found that the source spectrum would be affected by photon pile-up and as a result, the expected distributions were significantly discrepant from the observed ones. We therefore considered an annular region of 30" outer radius and different values of inner radii for the source and checked for the presence of pile-up. We found that, with an inner radius of 10", the source would be pile-up free\footnote{\url{https://www.cosmos.esa.int/web/xmm-newton/sas-thread-epatplot}}. We therefore used this inner radius for the spectral extraction. The response files (\textit{arf} and \textit{rmf} files) were generated by using {\tt SAS} tasks {\tt ARFGEN} and {\tt RMFGEN}, respectively. Background-subtracted light curves were produced using the {\tt FTOOLS} task {\tt LCMATH}. It is also to be noted that we ran SAS task `{\tt correctforpileup}' when we generated the pile-up corrected rmf file by using `{\tt rmfgen}'.

\label{sec:swift}
\subsection{\swift}

\swift~ monitored \source~ over a decade in both window-timing (WT) and photon counting (PC) modes. The source was observed simultaneously with \swift/XRT and \nustar~ three times (see Table~\ref{tab:log} for details). The $0.3-10$~keV spectra were generated using the standard online tools provided by the UK {\it Swift} Science Data Centre \citep{Evans2009} \footnote{\url{http://www.swift.ac.uk/user_objects/}}. For the present study, we used both WT and PC mode spectra in the $0.3-10$~keV range. We also generated long-term light curves in the $0.3-10$~keV band using the online-tools\footnote{\url{http://www.swift.ac.uk/user_objects/}}.

\begin{figure*}
\includegraphics[width=18cm]{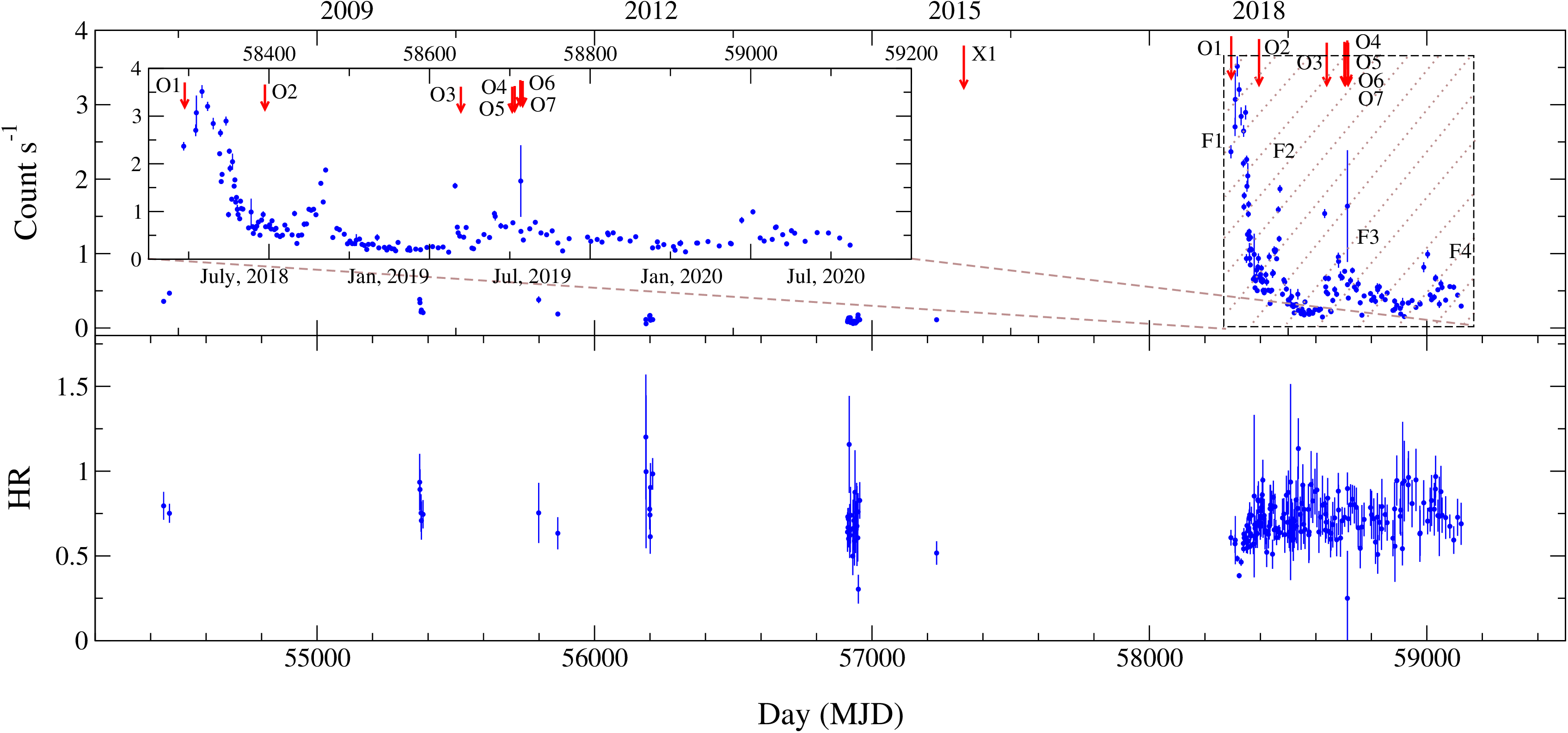} 
\caption{The long-term, $0.3-10$~keV \swift/XRT light curve of \source~ is shown in the top panel. The shaded region shows the light curve between 2018 and 2020. The inset figure in the top panel shows the expanded light curve of the shaded region from 2018 to 2020 for clarity. The red arrows represent the \nustar, \xmm~ and \swift/XRT observations of the source (see Table~\ref{tab:log} for details). In the bottom panel, we illustrate the variation of the hardness ratio (HR). The HR is defined as the ratio between count rates in the $1.5-10$~keV and the $0.3-1.5$~keV band.}
\label{fig:xrt-lc}
\end{figure*}

\label{sec:res}
\section{Results}

\begin{table*}
\caption{Variability in different energy bands are shown here. In some cases, the average error of observational data exceeds the 1$\sigma$ limit, resulting negative excess variance. In such cases, we obtained imaginary $F_{\rm var}$, which are not included in this table.}
\label{tab:var}
\begin{tabular}{lccccccccccccc}
\hline
ID & Energy & N & $F_{\rm max}$ & $F_{\rm min}$& R & Mean & $\sigma$        &$\sigma^2_{\rm XS}$&$\sigma^2_{\rm NXS}$ & $F_{\rm var}$ \\
 & (keV) & &(Count $s^{-1}$)&(Count $s^{-1}$) &  & &($10^{-3}$) & ($10^{-3}$) & ($10^{-3}$) & (\%)  \\
 (1) & (2) & (3) & (4) & (5) & (6) & (7) & (8) & (9) & (10) & (11)\\ 
\hline
X1& 0.5-3&  178& 1.04&  0.63 &  1.66&  0.82 &$ 71\pm2$&  2.74&$   3.3\pm0.7 $&$  5.8\pm0.6$\\
X1& 3-10 &  178& 0.19&  0.07 &  2.63&  0.12 &$ 19\pm1$& -- &$  -        $&$    -        $\\
\hline                                                                                             
O1& 0.5-3&  185&27.84&  16.69&  1.67&  21.89&$ 2573\pm78 $&  6.55&$   13.7\pm0.2 $&$ 11.7\pm0.6$\\
O1$^*$& 3-10 &  145& 2.35&  1.31 &  1.80&  1.81 &$ 210 \pm11 $&  2.2 &$   9.7\pm1.0 $&$  9.8\pm0.8$\\
O1& 10-70&  145& 1.21&  0.54 &  2.23&  0.80 &$ 98  \pm 5 $&  4.94&$   7.8\pm1.5 $&$  8.8\pm1.0$\\
\hline                                                                                             
O2& 0.5-3&  213& 5.69&  2.98 &  1.91&  4.05 &$ 628 \pm15 $& 380.2&$   93.9\pm0.6  $&$ 30.6\pm0.8$\\
O2$^*$& 3-10 &  197& 0.86&  0.29 &  2.99&  0.52 &$ 98\pm3  $&  6.50&$   24.4\pm2.6 $&$ 15.6\pm1.2$\\
O2& 10-70&  197& 0.47&  0.12 &  3.78&  0.25 &$ 51\pm2$&  1.11&$   17.2\pm0.4  $&$ 13.1\pm1.6$\\
\hline                                                                                             
O3& 0.5-3&  185& 3.74&  2.52 &  1.48&  3.13 &$ 289 \pm11 $&  72.2&$   23.1\pm0.5  $&$ 15.2\pm0.4$\\
O3$^*$& 3-10 &  161& 0.48&  0.08 &  6.46&  0.22 &$ 48\pm2$&  0.16&$   3.2\pm5.4 $&$  5.7\pm4.8$\\
O3& 10-70&  161& 0.47&  0.04 &  12.5&  0.16 &$ 45\pm2$&  0.28&$   11.2\pm0.9  $&$ 10.6\pm4.3$\\
\hline                                                                                                   
O4& 3-10 &  145& 0.99&  0.36 &  2.78&  0.55 &$ 86\pm3$&  4.76&$   15.4\pm2.2$&$ 12.4\pm1.1$\\
O4& 10-70&  145& 0.37&  0.15 &  2.42&  0.26 &$ 40\pm1$&  0.39&$    5.9\pm2.8$&$  7.7\pm1.8$\\
\hline                                                                                                   
O5& 0.5-3&  27 & 3.71&  3.33 &  1.12&  3.52 &$ 109 \pm12 $&  0.46&$    0.1\pm0.3$&$  0.6\pm2.2$\\
O5& 3-10 &  27 & 0.55&  0.15 &  3.68&  0.39 &$ 116 \pm3  $& 10.78&$   71.5\pm14.7$&$ 26.7\pm4.6$\\
\hline                                                                                             
O6& 3-10 &  191& 0.45&  0.16 &  2.77&  0.31 &$ 52\pm2$&  1.27&$   13.1\pm2.6$&$ 11.4\pm1.3$\\
O6& 10-70&  191& 0.23&  0.08 &  2.83&  0.16 &$ 30  \pm1  $&  0.12&$   4.9\pm3.8 $&$ 7.0\pm02.7$\\
\hline                                                                                                   
O7& 3-10 &  208& 1.46&  0.21 &  6.85&  0.35 &$ 98\pm7$&  2.50&$   20.3\pm7.4$&$ 14.3\pm2.7$\\
O7& 10-70&  208& 0.73&  0.09 &  8.22&  0.17 &$ 53\pm4$& --&$  -        $&$   -         $\\
\hline
\end{tabular}
\leftline{Columns in the table represent -- (1) ID of observation, (2) energy range, (3) number of data points or length of the light curve, (4) maximum count of }
\leftline{the light curve, (5) minimum count of the light curve, (6) ratio of maximum count to minimum count, $R=F_{\rm max}/F_{\rm min}$, (7) mean count of the light curve,} 
\leftline{(8) standard deviation of the light curve, (9) excess variance, (10) normalized excess variance, (11) fractional rms amplitude of the light curve.}
\leftline{$^{*}$During O1, O2 and O3, we reported only \nustar~ observation in $3-10$~keV energy range, although both \xmm~ and \nustar~ data}
\leftline{were available in $3-10$~keV energy band.}
\end{table*}

\begin{figure}
\includegraphics[width=8.5cm]{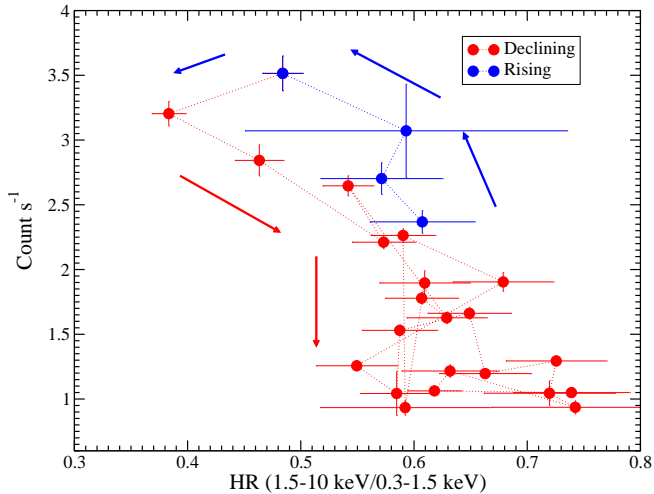}
\caption{Hardness-intensity diagram (HID) for the first outburst F1 (from MJD 58308 to MJD 58380). The $0.3-10$~keV \swift/XRT count rate is plotted as a function of hardness ratio (HR). The arrow marks in the figure represent the evolution of the outburst.}
\label{fig:hid}
\end{figure}

\label{sec:timing}
\subsection{Timing Analysis}

We studied the long term \swift/XRT light curves of \source~ in the $0.3-10$~keV energy range for the timing analysis. Along with the \swift/XRT light curves, the $0.5-10$~keV and $3-70$ keV light curves (500~s time-bin) from \xmm~ and \nustar~ observations were also analysed. 

\label{sec:prof}
\subsubsection{Outburst Profile}
\source~was observed intensively with many observatories for about 70 years, starting from the 1950s. In this time period, \source~ showed two major outbursts in optical wavebands, in 1962 and 1992, along with several flaring episodes \citep{Alloin1986,dasilva2017}. Before the major X-ray outburst in 2018, a flaring event was observed in 2010 by the \swift/BAT survey\footnote{\url{https://swift.gsfc.nasa.gov/results/bs105mon/216}} \citep{Oknyansky2018,Oknyansky2019}. Since then, \source~ remained in the low state with a luminosity of $L \sim 10^{41}$ \eps~ in 2--10 keV energy band.

\source~ was monitored by \swift/XRT over a decade, and it was caught in an outburst in June 2018, when the X-ray intensity increased by $\sim25-30$ times compared to the quiescent state \citep{Oknyansky2019}. During the 2018 X-ray outburst, the source also brightened in the optical, ultraviolet (UV) and infrared (IR) wavebands \citep{Dai18,Cutri18}. The optical and near-infrared (NIR) observations showed that the AGN started to brighten since 2017 September \citep{Dai18}. In the upper panel of Figure~\ref{fig:xrt-lc}, we show the long term $0.3-10$~keV \swift/XRT light curve. From this figure, it can be seen that the source experienced a major outburst in June 2018 (F1), which was followed by three smaller flaring events in December 2018 \citep[F2;][]{Grupe18b}, May 2019 \citep[F3;][]{Grupe19}, and May 2020 (F4). The smaller outbursts (F2, F3 and F4) were not as bright as the main outburst (F1). In the bottom panel of Figure~\ref{fig:xrt-lc}, we show the evolution of the hardness ratio (HR; i.e. the ratio between the $1.5-10$~keV and $0.3-1.5$~keV count rate) with time. Unlike the \swift/XRT light curve (top panel), the HR plot did not show any significant long-term variability. In Figure~\ref{fig:hid}, we show hardness-intensity diagram \citep{Homan2001,RM06} for the main outburst (F1), where the \swift/XRT count rate is plotted as a function of HR. The HID or `q'-diagram appeared to show a `q'-like shape, which is ubiquitous for outbursting Galactic black hole X-ray binaries. Interestingly, we did not observe any clear sign of `q'-shape HID for the next three recurrent outbursts.

\label{sec:variability}
\subsubsection{Variability}
We studied the source variability in different energy bands. As the soft excess in AGNs is generally observed below 3 keV, whereas the primary emission is observed above 3~keV, we analysed the \xmm~ light curves in $0.5-3$~keV and $3-10$~keV energy ranges. We also studied the \nustar~ light curves in two separate bands ($3-10$~keV and $10-70$~keV).

We calculated the peak-to-peak ratio of the light curves, which is defined as $R=F_{\rm max}/F_{\rm min}$, where $F_{\rm max}$ and $F_{\rm min}$ are the maximum and minimum count rates, respectively. The light curves in different energy bands ($0.5-3$~keV, $3-10$~keV, and $10-70$~keV) showed different magnitude of variability. In all observations, $R$ was higher in the $3-10$~keV energy band than in the $0.5-3$~keV range. In the $10-70$~keV energy band, $R$ was higher than in the lower energy band, except for O4. The mean value of $R$ in $0.5-3$~keV, $3-10$~keV, and $10-70$~keV energy bands are $<R> = 1.34, 3.43$, and $5.33$, respectively. Thus, it is clear that the light curves showed higher variability in the higher energy bands in terms of R. However, this is too simplistic as very high or low count could be generated due to systematic/instrumental error. Hence, we calculated the normalized variance ($\sigma_{\rm NXS}^2$) and fractional variability ($F_{\rm var}$ to study the variability.

We calculated the fractional variability ($F_{\rm var}$) \citep{Edelson1996,Edelson2001,Edelson2012,Nandra1997,Vaughan2003} in the 0.5--3, 3--10 and 10--70 keV bands for a light curve of $x_i$ counts $s^{-1}$ with uncertainties $\sigma_i$ of length $N$, with a mean $\mu$ and standard deviation $\sigma$ is given by,

\begin{equation}
F_{\rm var} = \sqrt{\frac{\sigma^2_{\rm XS}}{\mu^2}},
\end{equation}

where, $\sigma^2_{\rm XS}$ is the excess variance \citep{Nandra1997,Edelson2002} which is given by,

\begin{equation}
\sigma^2_{\rm XS}=\sigma^2 - \sigma^2_{\rm err},
\end{equation}

where, $\sigma^{2}_{\rm err}$ is the mean squared error. The $\sigma^{2}_{\rm err}$ is given by,

\begin{equation}
\sigma^2_{\rm err} =  \frac{1}{N}\sum_{i=1}^{N} \sigma^2_{\rm i}.
\end{equation}

The normalized excess variance is given by,

\begin{equation}
\sigma^2_{\rm NXS}=\frac{\sigma^2_{\rm XS}}{\mu^2}.
\end{equation}

The uncertainties in $F_{\rm var}$ and $\sigma_{\rm NXS}$ \citep{Vaughan2003} are given by,

\begin{equation}
{\rm err}(F_{\rm var})= \sqrt{(\sqrt{\frac{1}{2N}}\frac{\sigma_{\rm err}^2}{\mu^2 F_{\rm var}})^2+(\frac{1}{\mu}\sqrt{\frac{\sigma^2_{\rm err}}{N}})^2},
\end{equation}

and

\begin{equation}
{\rm err}(\sigma^2_{\rm NXS}) = \sqrt{(\sqrt{\frac{2}{N}}\frac{\sigma_{\rm err}^2}{\mu^2})^2+(\sqrt{\frac{\sigma_{\rm err}^2}{N}}\frac{2F_{\rm var}}{\mu})^2}.
\end{equation}

For a few observations, we could not estimate excess variance due to large errors in the data. Thus, from the normalized variance ($\sigma_{\rm NXS}^2$), the trend of variability was not clear. We calculated the fractional rms variability amplitude ($F_{\rm var}$) to study the variability. We obtained the highest fractional variability ($F_{\rm var}$) in the $0.5-3$~keV energy range for four observations (X1, O1, O2 \& O3), while the highest variability is observed in $3-10$~keV range for O5. The mean value of the fractional variability in $0.5-3$~keV, $3-10$~keV, and $10-70$~keV energy ranges were $<F_{\rm var}> = 12.8\pm 0.5\%, 12.0\pm 2.8\%$, and $7.9\pm 2.4 \%$, respectively. This indicates that the strongest variability is observed in the $0.5-3$~keV energy range. The variability parameters discussed here are reported in Table~\ref{tab:var}.

\label{sec:correlation}
\subsubsection{Correlation}

The soft excess is generally observed below 3~keV. To investigate the origin of the soft excess, we calculated the time delay between the soft X-ray photons ($0.5-3$~keV range) and the continuum photons ($3-10$~keV) using cross-correlation method from the \xmm~observations.
We used the $\xi$-transformed discrete correlation function (ZDCF) method \citep{Alexander1997}\footnote{\url{http://www.weizmann.ac.il/particle/tal/research-activities/software}} to investigate the time-delay between the soft-excess and the X-ray continuum. The ZDCF co-efficient was calculated for two cases: omitting the zero lag points and including the zero lag points. In both cases, similar results were obtained. A strong correlation between the $0.5-3$ keV and $3-10$ keV energy bands was observed (amplitude $>0.7$) during observations O1, O2 and O3. However, no significant delay was observed during those observations. The values of ZDCF coefficient with time delay are presented in Table~\ref{tab:lag}.

\begin{table}
\caption{ZDCF results.}
\label{tab:lag}
\begin{tabular}{lcccc}
\hline
  & soft excess& & & \\
ID & $\Delta t^{\dagger}$ & $\sigma^{\dagger}$ & amp$^{\dagger}$ \\  
 & (min) & (min) & \\
\hline
X1&  $-10.5\pm13.0$ & $145.2 \pm 8.3$& $0.24\pm0.07$  \\   
O1&  $13.8 \pm 3.1$ & $103.3\pm 8.3$& $0.84 \pm 0.08$ \\
O2&  $11.7 \pm 5.9$ & $222.9 \pm 8.3$& $0.90 \pm 0.07$  \\
O3&  $13.9 \pm 7.8$ & $263.4\pm 8.3$& $0.69 \pm 0.07$  \\
O5& -- & -- & -- \\
\hline
\end{tabular}
\leftline{$^{\dagger}$ ZDCF correlation between primary X-ray continuum ($3-10$ energy band) }
\leftline{and soft excess ($0.5-3$~keV energy band). $\sigma$'s and amp's are FWHM and }
\leftline{amplitude of the ZDCF function. Note that the maximum amplitude can be 1.}
\end{table}

\begin{figure*}
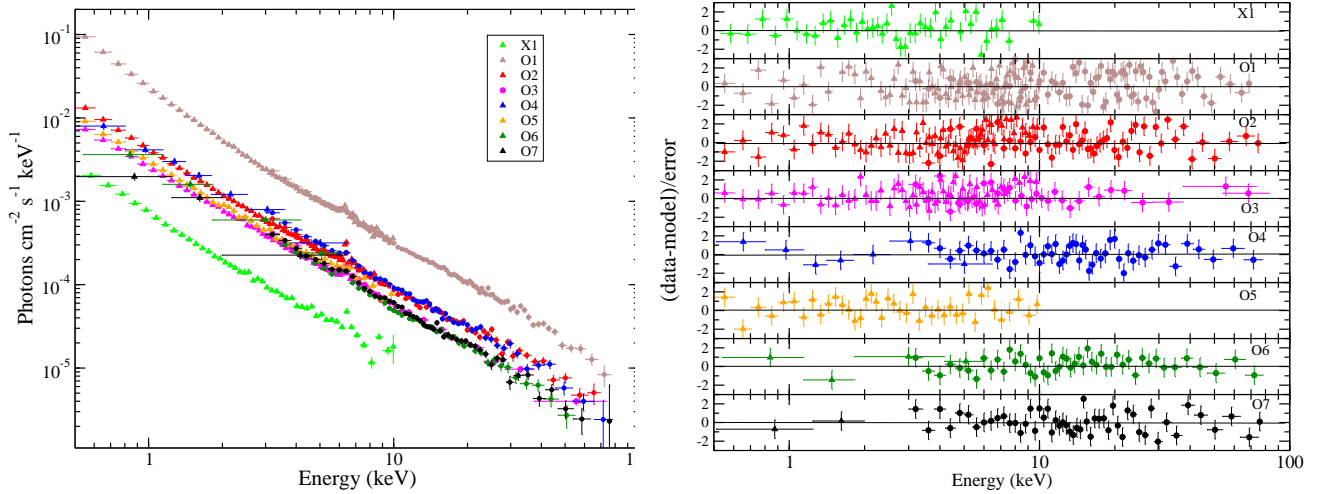

\includegraphics[width=8.5cm]{spec-1.eps}
\includegraphics[width=8.5cm]{del-1.eps}
\caption{The left panel shows the unfolded spectra obtained from each observations. Triangles represent the \xmm~ or \swift/XRT data, while circles represent the \nustar~ data. The light-green, brown, red, magenta, blue, orange, dark green and black points represent the observation X1, O1, O2, O3, O4, O5, O6 and O7, respectively. The residuals obtained after fitting the source spectra with Model--1 and Model--3 are shown in the right panels. The spectra are re-binned for visual purpose.}
\label{fig:spec1}
\end{figure*}

\label{sec:spec}
\subsection{Spectral Analysis}
We carry out the X-ray spectral analysis using data obtained from the {\it Swift}/XRT, \xmm~ and \nustar~ observations of \source~  using {\tt XSPEC} v12.10 package  \citep{Arnaud1996}\footnote{\url{https://heasarc.gsfc.nasa.gov/xanadu/xspec/}}. The spectral analysis was performed using simultaneous \xmm~ and \nustar~ observations in the $0.5-70$~keV energy band for three epochs, simultaneous \swift/XRT and \nustar~ observations ($0.5-70$~keV) for three epochs, and \xmm~ observations for two epochs ($0.5-10$~keV), between 2015 November 5 and 2019 August 21 (see Table~\ref{tab:log}).

For the spectral analysis, we used various phenomenological and physical models, namely, {\tt powerlaw} (PL)\footnote{\url{https://heasarc.gsfc.nasa.gov/xanadu/xspec/manual/node213.html}}, {\tt NTHCOMP}\footnote{\url{https://heasarc.gsfc.nasa.gov/xanadu/xspec/manual/node205.html}} \citep{Z96,Zycki1999}, {\tt OPTXAGNF}\footnote{\url{https://heasarc.gsfc.nasa.gov/xanadu/xspec/manual/node206.html}} \citep{Done2012}, and {\tt RELXILL}\footnote{\url{www.sternwarte.uni-erlangen.de/~dauser/research/relxill/}} \citep{Garcia2014,Dauser2014} to understand the X-ray properties of NGC~1566. In general, an X-ray spectrum of an AGN consists of a power-law continuum, a reflection hump at around $15-40$~keV, a Fe K$\alpha$ fluorescent line, and a soft X-ray component below 2 keV \citep{Netzer2015,Padovani2017,Ricci2017}. The observed X-ray emission also suffers from absorption caused by the interstellar medium and the torus. In our analysis, we used two absorption components. For the Galactic interstellar absorption, we used {\tt TBabs}\footnote{\url{https://heasarc.gsfc.nasa.gov/xanadu/xspec/manual/node265.html}}\citep{Wilms2000} with fixed hydrogen column density of $N_{\rm H} = 7.15 \times 10^{19}$ \pcm \citep{HI4PI2016}. In addition, we also used a ionized absorption model {\tt zxipcf}. For both the absorption components, we used the {\it wilms} abundances \citep{Wilms2000} and the Verner cross-section \citep{Verner1996}. In this work, we used the following cosmological parameters : $H_0= 70$ km s$^{-1}$ Mpc $^{-1}$, $\Lambda_0 = 70$, and $\sigma_M= 0.27$ \citep{Bennett2003}. The uncertainties in each spectral parameters are calculated using the `{\tt error}' command in {\tt XSPEC}, and reported at the 90\% confidence (1.6 $\sigma$).

\begin{figure*}
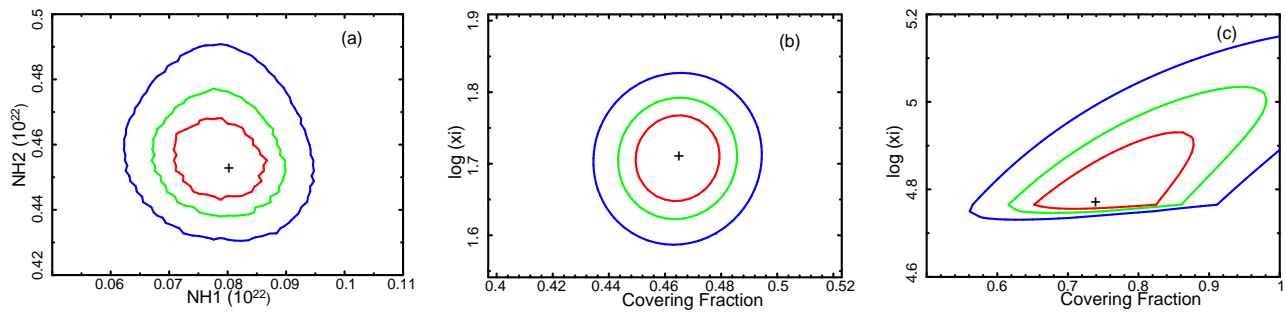

\includegraphics[angle=270,width=6cm]{nh-nh.eps} \hskip-0.3cm
\includegraphics[angle=270,width=6cm]{xi-cf-1.eps} \hskip-0.3cm
\includegraphics[angle=270,width=6cm]{xi-cf-2.eps}
\caption{Left panel shows the 2D contour between the column density of the low-ionizing absorber ($N_{\rm H,1}$) and high-ionizing absorber ($N_{\rm H,2}$) for O1. The middle and right panel show 2D contour between log $\xi$ and covering fraction for low and high-ionizing absorber, respectively.}
\label{fig:cntr}
\end{figure*}

\begin{figure}
\centering
\includegraphics[width=8.5cm]{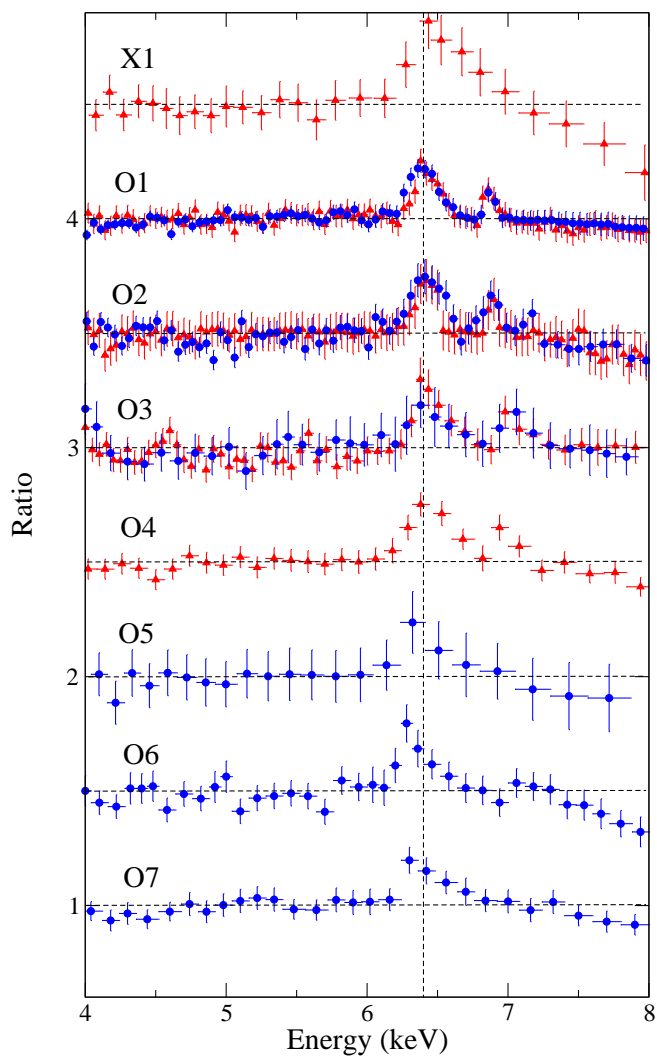}
\caption{Evolution of Fe complex. The ratio of \xmm~ and \nustar~ data to the power-law continuum model for every  observations are shown. The red triangles and blue circles represent the ratio obtained from \xmm~ and \nustar~ data, respectively. The vertical dashed line represent E=6.4~keV. The horizontal dashed lines represent the data/model=1 for each observation. The ratios are re-scaled by adding 0.5 in the y-axis separating the observations.}
\label{fig:fe}
\end{figure}

\subsubsection{Model 1: POWERLAW}
We built our baseline spectral model with {\tt power-law} continuum, along with the soft-excess, and Fe K-line emission. We used a blackbody component ({\tt bbody} in {\tt XSPEC}) for the soft-excess, and one or more Gaussian functions to incorporate the Fe~K complex. Out of the eight epochs, two Gaussian lines were required for six observations. Two ionized absorbers were also needed while fitting the data from three observations. The final model (hereafter Model--1) reads in {\tt XSPEC} as, 

{\tt TB $\times$ zxipcf1 $\times$ zxipcf2 $\times$ ( zPL1+ zGa + zGa + bbody)}.

We started our analysis with the pre-outburst \xmm~ observation X1 (in the $0.5-10$~keV energy range), $\sim$2.5 years prior to the 2018 outburst. Model--1 provided a good fit to the \xmm~ data, with $N_{\rm H} = (3.53 \pm 0.06) \times 10^{21}$ \pcm~ and power-law photon index of $\Gamma = 1.72\pm 0.05$, with $\chi^2 = 1073$ for 998 degrees of freedom (dof). Along with this, an iron K$\alpha$ emission line at $\sim$6.4~keV with an equivalent width (EW) of $\sim 206^{+4}_{-18}$ eV was detected.

Next, we analyzed simultaneous observations of \source~ with \xmm~ and \nustar~ in the rising phase of the 2018 outburst (O1). First, we included one {\tt zxipcf} component in our spectral model. The fit returned with $\chi^2=2850$ for 2562 degrees of freedom (dof). However, negative residuals were clearly observed in soft X-rays  (<1 keV). Thus, we included another {\tt zxipcf} component, and our fit improved significantly with $\Delta \chi^2=106$ for 3 dof. The spectral fitting in $0.5-70$~keV range returned $\Gamma = 1.85\pm0.04$, and $\chi^2 = 2744$ for 2559 dof. The Fe~K$\alpha$ line was detected at 6.38 keV with EW of $114\pm15$~eV, as well as another emission feature at 6.87 keV, with EW $<37$ eV. This line could be associated with Fe~XXVI line. We required two ionized absorber to fit the spectra, one low-ionization absorber ($\xi \sim 10^{1.7 \pm 0.1}$) with $N_{\rm H,1} = (8.1 \pm 2.2) \times 10^{20}$ \pcm, and one high-ionization absorber ($\xi \sim 10^{4.7 \pm 0.4}$) with $N_{\rm H,2} = (4.3 \pm 0.4) \times 10^{21}$ \pcm. The high-ionizing absorber required a high covering fraction ($CF>0.73$), while the low-ionizing absorber a moderate covering fraction with $CF \sim 0.46 \pm 0.04$. 

The next observation (O2) was carried out on 2018 October 10, simultaneously with \xmm~ and \nustar, covering the $0.5-70$~keV energy range. The source was in the decay phase of the outburst at the time of the observation. We started our fitting with one {\tt zxipcf} component. Although, the model provided a good fit to the data with $\chi^2=2298$ for 2130 dof, an absorption feature was seen in the residuals. Thus, we added a second {\tt zxipcf} component, and our fit returned with $\chi^2$ = 2224 for 2127 dof. The photon index decreased marginally compared to O1 ($\Gamma = 1.78\pm0.02$), while the column density increased slightly for both low and high-ionizing absorbers. The Fe~K$\alpha$ and Fe~XXVI lines were detected at 6.41 keV and 6.89 keV, with EWs of $126^{+3}_{-21}$~eV and $<49$~eV, respectively. The next simultaneous observations of \source~ with \xmm~ and \nustar~ (O3) were carried out $\sim$8 months after the end of the outburst. The source was in a low state during the observation. Similar to O1 and O2, adding second absorption component improved the fit $\Delta \chi^2=67$ for 3 dof. The column density increased to $N_{\rm H,1} = (1.24 \pm 0.14) \times 10^{21}$ \pcm~ for the low-ionization absorber, while the column density decreased to $N_{\rm H,2} = (8.9\pm0.2) \times 10^{20}$ \pcm~ for the high-ionization absorber. The photon index was found to be $\Gamma = 1.68\pm 0.02$ in this observation. We detected both Fe~K$\alpha$ and Fe~k$\beta$ lines at 6.39~keV and 7.04 keV, with EWs of $117\pm 14$~eV and $<92$~eV, respectively. 

\begin{figure*}
\centering
\includegraphics[width=18cm]{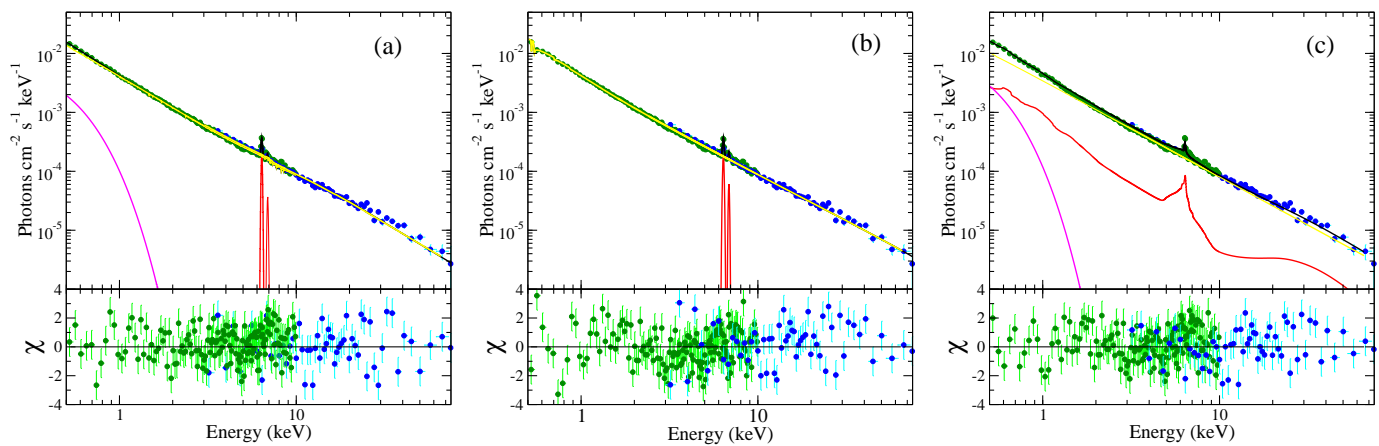}
\caption{Best-fit unfolded spectra using Model--1, Model--3 and Model--4 are shown in the left, middle, and right panel, respectively, for O2. The corresponding residuals are shown in the bottom of each panel. Left panel: the black, yellow, magenta and red lines represent the total, primary emission, soft excess and Fe line emission, respectively. Middle panel: the black, yellow, and red lines represent the total, the AGN emission and Fe line emission, respectively. Right panel: the black, yellow, magenta and red lines represent the total, primary emission, soft excess and reflection component, respectively.}
\label{fig:uf-spec}
\end{figure*}

The last four observations (O4, O5, O6 \& O7) were carried out in the span of 13 days. Observation O4 was made during the second small flare (F2), after the 2018 main outburst. During these four observations, the photon index was nearly constant ($\Gamma \sim 1.67\pm0.07-1.69\pm0.06$), while the column density of the low-ionizing absorber was found to vary in the range of $N_{\rm H,1} \sim 1.2-1.3 \times 10^{21}$ \pcm. A high-ionization absorber was not required to fit the spectra of these four observations. We also observed a low covering fraction in these four observations (see Table~\ref{tab:mod1}). The Fe K$\alpha$ line was detected in all four observations, with EW $> 100$ eV (except O5). During our observations, the blackbody temperature ($T_{\rm bb}$) was observed to be remarkably constant with $T_{\rm bb} \sim 110$~eV. The parameters obtained by our spectral fitting results are listed in Table~\ref{tab:mod1}. Model--1 fitted spectra of \source~ are shown in the left panel of Fig.~\ref{fig:spec1}, whereas the corresponding residuals are shown in the right panels. To test for the presence of degeneracies between the column densities of two ionizing absorbers, we plotted the 2D contour in Fig.~\ref{fig:cntr}a for the observation O1. In Fig.~\ref{fig:cntr}b and Fig.~\ref{fig:cntr}c, we show 2D contours between log($\xi$) and covering fraction for the low and high-ionizing absorbers, respectively, for observation O1. In Fig~\ref{fig:fe}, we show the residuals above the continuum in $4-8$~keV energy range for Fe line emission. In Fig.~\ref{fig:uf-spec}a, we also show the unfolded spectrum fitted with Model--1 for observation O2.

\begin{table*}
\caption{Best-fit parameters obtained from the spectral fitting of the source spectra with Model--1 ({\tt POWER-LAW}) \& Model--2 ({\tt NTHCOMP}).}
\label{tab:mod1}
\hspace*{-0.5in}
\begin{tabular}{lccccccccccccccc}
\hline
& X1 & O1 & O2 & O3 & O4 & O5 & O6 & O7\\
\hline
$N_{\rm H,1}$ ($10^{21}$ \pcm) & $3.53^{+0.05}_{-0.06}$ & $0.81^{+0.13}_{-0.22}$ & $0.96^{+0.15}_{-0.18}$ & $1.24^{+0.14}_{-0.11}$& $1.33^{+0.14}_{-0.10}$ & $1.18^{+0.08}_{-0.09}$& $1.25^{+0.12}_{-0.16}$ & $1.30^{+0.17}_{-0.22}$ \\
\\
log${\rm \xi_{\rm 1}}$ & $-3^{\dagger}$ & $1.71^{+0.12}_{-0.11}$ & $1.81^{+0.10}_{-0.08}$ & $1.37^{+0.08}_{-0.09}$ & $1.10^{+0.04}_{-0.08}$& $0.26^{+0.09}_{-0.05}$& $0.17^{+0.14}_{-0.05}$& $0.21^{+0.03}_{-0.08}$ \\
\\
Cov Frac1 & $0.20^{+0.13}_{-0.05}$ & $0.46^{+0.04}_{-0.03}$ & $0.31^{+0.09}_{-0.04}$ & $0.33^{+0.07}_{-0.12}$ & $<0.12$ & $0.24^{+0.12}_{-0.15}$ & $0.17^{+0.04}_{-0.08}$& $<0.1$ \\
\\
$N_{\rm H,2}$ ($10^{21}$ \pcm) & -- & $4.31^{+0.41}_{-0.26}$ & $4.56^{+0.58}_{-0.47}$ &$0.89^{+0.22}_{-0.17}$ & -- & -- & -- & --\\
\\
log${\rm \xi_{\rm 2}}$ & -- & $4.73^{+0.40}_{-0.02}$ & $3.56^{+0.26}_{-0.08}$ &  $3.07^{+0.19}_{-0.07}$& -- & -- & --& -- \\
\\
Cov Frac2 & -- &$>0.73$ & $0.61^{+0.26}_{-0.21}$ & $0.54^{+0.23}_{-0.39}$ & -- & --&-- & -- \\
\\
$\Gamma$&$ 1.72^{+0.05}_{-0.05}$  &$ 1.85^{+0.04}_{-0.04}$  &$ 1.78^{+0.02}_{-0.02}$  &$ 1.68^{+0.02}_{-0.02}$  &$ 1.67^{+0.05}_{-0.07}$  &$ 1.68^{+0.03}_{-0.04}$  &$ 1.69^{+0.04}_{-0.06}$ &$ 1.67^{+0.05}_{-0.07}$ \\
\\
PL Norm ($10^{-3}$)&$ 1.59^{+0.16}_{-0.18}$  &$14.6^{+1.43}_{-2.05}$  &$ 5.97^{+0.06}_{-0.10}$  &$ 2.45^{+0.05}_{-0.10}$  &$ 4.71^{+0.07}_{-0.08}$  &$ 2.84^{+0.13}_{-0.09}$  &$ 2.78^{+0.11}_{-0.15}$ &$ 2.58^{+0.10}_{-0.17}$ \\
\\
Fe K$\alpha$ ~~~~~~LE (keV)&$ 6.44^{+0.03}_{-0.04}$  &$ 6.38^{+0.04}_{-0.04}$  &$ 6.41^{+0.03}_{-0.06}$  &$ 6.39^{+0.07}_{-0.10}$  &$ 6.29^{+0.08}_{-0.14}$  &$ 6.28^{+0.07}_{-0.11}$  &$ 6.29^{+0.05}_{-0.09}$ &$ 6.19^{+0.05}_{-0.10}$ \\
\\
~~~~~~~~~~~~~~~~~~EW (eV)&$ 206^{+4}_{-18}$  &$ 114^{+11}_{-15}$ &$ 126^{+3}_{-21}$ &$117^{+14}_{-10 }$ &$ 106^{+1}_{-6}$  &$ <95$  &$ 155^{+12 }_{-16 }$ &$  108^{+15 }_{-22}$ \\
\\
~~~~~~~~~~~~~~~~~~FWHM (km s$^{-1}$)&$ 2329^*$  &$ 8695^{+924 }_{-1243 }$  &$ 2108^*$  &$ 4796^{+863 }_{-982 }$  &$ 2337^*$  &$ 4461^{+1045}_{-845 }$  &$ 6025^{+946 }_{-1194 }$ &$ 6613^{+1275 }_{-1223 }$ \\
\\
~~~~~~~~~~~~~~~~~~Norm ($10^{-5}$)&$ 6.83^{+0.15}_{-0.22}$  &$11.74^{+1.10}_{-1.22}$  &$ 3.47^{+0.14}_{-0.10}$  &$ 2.56^{+0.18}_{-0.13}$  &$ 1.91^{+0.07}_{-0.09}$  &$ 0.93^{+0.04}_{-0.10}$  &$ 2.14^{+0.10}_{-0.13}$ &$ 2.43^{+0.10}_{-0.08}$ \\
\\
Fe XXVI ~~LE (keV)&$ -$  &$ 6.87^{+0.04}_{-0.04}$  &$ 6.89^{+0.04}_{-0.03}$  &$ 7.04^{+0.05}_{-0.04}$  &$ 6.93^{+0.04}_{-0.03}$  &$ -                   $  &-- &-- \\
\\
~~~~~~~~~~~~~~~~~~EW (eV)&$ - $  &$   <37$  &$   <49$  &$   <92$  &$   <42$  &$-  $  & -- &$   -$ \\
\\
~~~~~~~~~~~~~~~~~~FWHM (km s$^{-1}$) &$ -                   $  &$ <4452$  &$ <4049$  &$ <4943$  &$ <3913$  &$ -     $  & -- & -- \\
\\
~~~~~~~~~~~~~~~~~~Norm ($10^{-6}$)&$- $  &$21.09^{+1.02}_{-0.94}$  &$ 7.88^{+0.13}_{-0.22}$  &$ 8.95^{+1.06}_{-0.94}$  &$ 7.30^{+0.31}_{-0.73}$  &$ -                   $  & -- & -- \\    
\\
$kT_{\rm bb1}$ (eV) & $116^{+7}_{-8}$ & $112^{+6}_{-8}$ &$117^{+11}_{-7}$ & $115^{+5}_{-9}$& $108^{+8}_{-12}$& $114^{+9}_{-6}$ & $117^{+12}_{-14}$& $122^{+17}_{-10}$ \\
\\
$N_{\rm bb1}$ ($10^{-5}$) &$0.49^{+0.03}_{-0.06}$ & $28.6^{+2.29}_{-3.42}$&$8.30^{+1.09}_{-0.65}$ & $4.35^{+0.76}_{-0.96}$& $5.33^{+0.51}_{-1.04}$& $0.98^{+0.24}_{-0.19}$& $4.71^{+0.72}_{-1.13}$& $2.65^{+0.80}_{-0.56}$\\
\\
$\chi^2$/dof&1073/998 &2744/2559 &2224/2127   &1608/1646    & 693/654 & 835/834 & 535/542  & 576/567    \\
\hline
$N_{\rm H,1}$ ($10^{21}$ \pcm) & $3.46^{+0.06}_{-0.05}$ & $0.78^{+0.10}_{-0.21}$ & $0.96^{+0.17}_{-0.16}$ & $1.22^{+0.13}_{-0.15}$& $1.32^{+0.12}_{-0.16}$ & $1.19^{+0.10}_{-0.09}$& $1.26^{+0.12}_{-0.18}$ & $1.27^{+0.15}_{-0.19}$ \\
\\
log${\rm \xi_{\rm 1}}$ & $-3^{\dagger}$ & $1.66^{+0.10}_{-0.08}$ & $1.85^{+0.06}_{-0.11}$ & $1.32^{+0.07}_{-0.10}$ & $1.05^{+0.04}_{-0.06}$& $0.26^{+0.05}_{-0.07}$& $0.15^{+0.12}_{-0.07}$& $0.21^{+0.04}_{-0.05}$ \\
\\
Cov Frac1 & $0.20^{+0.07}_{-0.05}$ & $0.46^{+0.03}_{-0.03}$ & $0.32^{+0.07}_{-0.04}$ & $0.34^{+0.05}_{-0.14}$ & $<0.12$ & $0.22^{+0.11}_{-0.16}$ & $<0.15$& $<0.1$ \\
\\
$N_{\rm H,2}$ ($10^{21}$ \pcm) & -- & $4.25^{+0.32}_{-0.23}$ & $4.50^{+0.64}_{-0.43}$ &$0.82^{+0.27}_{-0.19}$ & -- & -- & -- & --\\
\\
log${\rm \xi_{\rm 2}}$ & -- & $4.63^{+0.45}_{-0.09}$ & $3.32^{+0.33}_{-0.10}$ &  $3.11^{+0.23}_{-0.16}$& -- & -- & --& -- \\
\\
Cov Frac2 & -- &$>0.75$ & $0.59^{+0.22}_{-0.25}$ & $0.56^{+0.23}_{-0.45}$ & -- & --&-- & -- \\
\\
$\Gamma$&$1.74 ^{+0.05}_{-0.07}$  &$ 1.88^{+0.04}_{-0.06}$  &$ 1.75^{+0.03}_{-0.03}$  &$ 1.72^{+0.02}_{-0.03}$  &$ 1.71^{+0.03}_{-0.05}$  &$ 1.69^{+0.04}_{-0.04}$  &$ 1.68^{+0.05}_{-0.05}$ &$ 1.69^{+0.04}_{-0.05}$ \\
\\
$kT_{\rm e}$ (keV)  &$101.8^{+4.9}_{-4.3}$  &$60.8^{+5.5}_{-6.7}$  &$85.7^{+7.4}_{-5.9}$  &$92.2^{+8.4}_{-9.4}$  &$75.2^{+8.3}_{-7.6}$  &$104.2^{+5.6}_{-6.9}$  &$105.8^{+6.9}_{-5.5}$ &$96.7^{+7.5}_{-8.2}$ \\
\\
$\tau$ & $1.27^{+0.17}_{-0.11}$  &$1.60^{+0.33}_{-0.18}$  &$1.43^{+0.16}_{-0.15}$  &$1.41^{+0.17}_{-0.12}$  &$1.66^{+0.24}_{-0.17}$  &$1.30^{+0.14}_{-0.12}$  &$1.34^{+0.15}_{-0.13}$ &$1.41^{+0.19}_{-0.13}$ \\
\hline
\end{tabular}
\leftline{$\xi$'s are in the unit ergs cm s$^{-1}$. PL norm is in unit of \phc. Fe line norm are in the unit of \phc.}
\leftline{$\tau$'s are calculated using Eqn. 1.}
\leftline{$\dagger$ pegged at the lowest value.}
\leftline{$^*$ Gaussian fitted with fixed line width, $\sigma=0.05$~keV.}
\end{table*}

\subsubsection{Model 2: NTHCOMP}
While fitting the source spectra with Model--1 provided us with information on the spectral shape and hydrogen column density, it provided limited information on the physical properties of the Comptonizing plasma. The Comptonizing plasma can be characterized by the electron temperature ($kT_{\rm e}$) and optical depth ($\tau$). In order to understand the properties of the Compton cloud, we replaced the {\tt powerlaw} continuum model with {\tt NTHCOMP} \citep{Z96,Zycki1999} in Model--1. The {\tt NTHCOMP} model provided us the photon index ($\Gamma$) and the hot electron temperature of the Compton cloud ($kT_{\rm e}$). The optical depth can be easily calculated from the information on $\Gamma$ and $kT_{\rm e}$ using the following equation \citep{ST80,Z96},

\begin{equation}
\tau \sim \sqrt{\frac{9}{4}+\frac{m_e c^2}{kT_{\rm e}}\frac{3}{(\Gamma-1)(\Gamma+2)}} -\frac{3}{2}.
\label{eqn:tau}
\end{equation}

This model (hereafter Model--2) reads in {\tt XSPEC} as,

{\tt TB $\times$ zxipcf1 $\times$ zxipcf2 $\times$ (NTHCOMP + zGa + zGa + bbody)}.

We fixed the seed photon temperature at $kT_{\rm s} = 30$~eV, which is reasonable for a BH of mass $\sim 8.3 \times 10^6$ \M \citep{SS1973,Makishima2000}. We required two absorption component during O1, O2 and O3. Inclusion of second absorption improved the fit significantly with $\Delta \chi^2=$~108, 84, 78 for 3 dof, during O1, O2 and O3, respectively. The photon indices and column densities obtained are similar to those obtained using Model--1. We found that $kT_{\rm e} = 102 \pm 5$~keV for observation X1. During the rising phase of the outburst (observation~O1), the Compton cloud was found to be relatively cooler, with $kT_{\rm e} = 61\pm 7$~keV. The Compton cloud was hot during the later observations, with the electron temperature increasing up to $106\pm 7$~keV within the eight observations analyzed here. A nearly constant photon index and a variable temperature would imply a variation in the density of the Comptonizing cloud. This would suggest that the optical depth was varying during the observations in the range of $\sim 1.27-1.66$. The results obtained from our spectral fitting with this model are listed in Table~\ref{tab:mod1}. 

\begin{figure}
\centering
\includegraphics[angle=270,width=8.5cm]{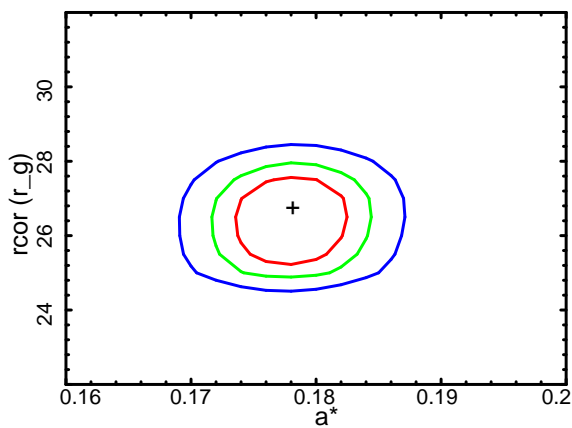}
\caption{2D contour plot between $R_{\rm cor}$ and $a^*$ for Model--3 during the observation O2.}
\label{fig:rcor-a}
\end{figure}

\subsubsection{Model 3: OPTXAGNF}

The X-ray spectra of AGN typically show a soft excess below 2 keV \citep{Arnaud1985,Singh1985}. Although the soft excess in AGNs was first detected in the 1980s, its origin is still very debated. In models 1 \& 2, we fitted the soft excess component with a phenomenological {\tt blackbody} model. To shed light on the origin of the soft excess in this source, we used a more physical model, {\tt OPTXAGNF} \citep{Done2012}.

The {\tt OPTXAGNF} model (hereafter Model--3) computes the soft excess and the primary emission self-consistently. In this model, total emission is determined by the mass accretion rate and by the BH mass. The disc emission emerges as a colour temperature corrected blackbody emission at radii $R_{\rm out} > r > R_{\rm cor}$, where $R_{\rm out}$ and $R_{\rm cor}$ are the outer edge of the disc and the corona, respectively. At $r<R_{\rm cor}$, the disc emission emerges as the Comptonized emission from a warm and optically-thick medium, rather than thermal emission. The hot and optically-thin corona is located around the disc and produces the high energy power-law continuum. The total Comptonized emission is divided between the cold and hot corona, and the fraction of the hot-Comptonized emission ($f_{\rm PL}$) can be found from the model fitting. The temperature of the cold corona ($kT_{\rm S}$), temperature of the seed photon, and the optical depth of the cold corona ($\tau$) at $r=R_{\rm cor}$ determine the energy of the up-scattered soft excess emission. The power-law continuum is approximated as the {\tt NTHCOMP} model, with the seed photon temperature fixed at the disc temperature at $r=R_{\rm cor}$, and the electron temperature fixed at 100 keV. 

When using the {\tt OPTXAGNF} model, we included two Gaussian components to incorporate the Fe emission lines. The model reads in {\tt XSPEC} as,

{\tt TB $\times$ zxipcf1 $\times$ zxipcf2 $\times$ ( OPTXAGNF + zGa + zGa)}.

While fitting the data with this model, we kept the BH mass frozen at $M_{\rm BH} = 8.3 \times 10^6$ \M~ \citep{Woo2002}. As recommended, we fixed normalization to unity during our analysis. Initially, we started our analysis with one absorption component. However, an absorption feature was seen during O1, O2 and O3. Thus, we added second absorption component in the spectra of O1, O2 and O3. Adding second {\tt zxipcf} component significantly improved the fit with $\Delta \chi^2 =$~88, 76, 65 for 3 dof, during the observation O1, O2 and O3, respectively. Overall, this model provided a good fit for all the observations. A clear variation in the Eddington ratio and size of the corona were observed in the different observations. In the rising phase of the 2018 outburst (observation O1), a high Eddington ratio ($L/L_{\rm Edd} \sim 0.23$) and a large corona ($R_{\rm cor} = 43\pm 3~R_{\rm g}$) were observed. These values were found to be higher than the pre-outburst observation (X1; $L/L_{\rm Edd} \sim 0.04$; $R_{\rm cor} = 12^{+2}_{-1}$ $R_{\rm g}$). In later observations, both Eddington ratio and size of the X-ray corona decreased to $L/L_{\rm Edd} \sim 0.06-0.07$ and $R_{\rm cor} \sim 15\pm 2-26\pm 2$ $R_g$, respectively. In the observation~O1, the electron temperature of the optically thick Comptonizing region was observed to be $kT_{\rm S} \sim  1.4 \pm 0.1$~keV, along with optical depth $\tau \sim 4.6 \pm 1$. The temperature of the optically-thick medium decreased to $kT_{\rm S} \sim 0.5-0.6$~keV in the later observations (O3 -- O7). However, the optical depth did not change much and varied in the range of $\tau \sim 4-5$. A significant fraction of the Comptonized emission was emitted from the optically thin corona with $f_{\rm PL} > 0.79$ during all the observations. We allowed the spin of the BH to vary. The best-fitted spin parameter fluctuated in the range of $a^* \sim 0.18^{+0.01}_{-0.02} - 0.21^{+0.03}_{-0.02}$, suggesting that the SMBH is spinning slowly. The variation of the column density ($N_{\rm H}$) and of the photon indices ($\Gamma$) was similar to that observed using Model--1. All the parameters obtained from our spectral analysis using Model--3 are presented in Table~\ref{tab:opt}. In Fig.~\ref{fig:uf-spec}b, we show the unfolded spectrum fitted with Model--3 for observation O2. In Fig.~\ref{fig:rcor-a}, we display the contour plot of $R_{\rm cor}$ and $a^*$, which shows that there is no strong degeneracy between those two parameters.

\begin{table*}
\caption{Best-fit parameters obtained from the spectral fitting of the source spectra with Model--3 ({\tt OPTXAGNF}).}
\label{tab:opt}
\begin{tabular}{lccccccccccccc}
\hline
&X1                      &O1                       &O2                     & O3                     &  O4                     &O5                       & O6                      &O7   \\
\hline                     
$N_{\rm H,1}$ ($10^{21}$ \pcm) & $3.49^{+0.05}_{-0.05}$  & $0.80^{+0.08}_{-0.16}$  &$0.94^{+0.15}_{-0.15}$ &$1.20^{+0.15}_{-0.11}$  &$1.34^{+0.10}_{-0.12}$   &$1.17^{+0.08}_{-0.11}$   &$1.25^{+0.08}_{-0.14}$   &$1.29^{+0.15}_{-0.22}$\\
\\
log${\rm \xi_{\rm 1}}$ &$-3^{\dagger}$          & $1.67^{+0.07}_{-0.06}$  &$1.82^{+0.03}_{-0.06}$ &$1.30^{+0.07}_{-0.09}$  &$1.08^{+0.07}_{-0.10}$   &$0.23^{+0.07}_{-0.10}$   &$0.18^{+0.06}_{-0.04}$   &$0.20^{+0.03}_{-0.04}$\\
\\
Cov Frac2 &$0.18^{+0.08}_{-0.06}$  & $0.45^{+0.03}_{-0.04}$  &$0.34^{+0.09}_{-0.06}$ &$0.35^{+0.05}_{-0.11}$  &$<0.12$                  &$0.21^{+0.12}_{-0.17}$   &$<0.14$& $<0.15$ \\
\\
$N_{\rm H,2}$ ($10^{21}$ \pcm) & --                      & $4.28^{+0.45}_{-0.38}$  &$4.56^{+0.51}_{-0.46}$ &$0.89^{+0.21}_{-0.16}$  & --                      & -- & -- & --\\
\\
log${\rm \xi_{\rm 2}}$ &--                      & $4.52^{+0.39}_{-0.22}$  &$3.15^{+0.21}_{-0.18}$ &$2.93^{+0.32}_{-0.19}$  & -- & -- & --& -- \\
\\
Cov Frac2 &--                      &$>0.79$                  &$0.54^{+0.22}_{-0.17}$ &$0.51^{+0.28}_{-0.21}$  & -- & --&-- & -- \\
\\
$L/L_{\rm Edd}$ &$-1.98^{+0.03}_{-0.04}$ &$-0.64^{+0.02}_{-0.02}$  &$-0.78^{+0.03}_{-0.03}$&$-1.16^{+0.03}_{-0.04}$ &$-0.81^{+0.02}_{-0.03}$  &$-0.97^{+0.04}_{-0.05}$  & $-1.20^{+0.03}_{-0.03}$ &$-1.17^{+0.02}_{-0.01}$\\
\\
$a^*$& $0.19^{+0.01}_{-0.01}$  &$0.18^{+0.01}_{-0.02}$   &$0.18^{+0.01}_{-0.01}$ &$0.21^{+0.01}_{-0.02}$  &$0.20^{+0.02}_{-0.03}$   &$0.19^{+0.03}_{-0.02}$   & $0.21^{+0.03}_{-0.02}$  &$0.19^{+0.01}_{-0.01}$\\
\\
$R_{\rm cor}$ ($R_{\rm g}$) &$12^{+2}_{-1}$          &$43^{+3}_{-2}$           &$26^{+2}_{-2}$         &$15^{+1}_{-2}$          &$22^{+2}_{-2}$           &$20^{+2}_{-3}$           & $17^{+2}_{-3}$          &$18^{+1}_{-1}$\\
\\
$kT_{\rm S}$ (keV) &$0.32^{+0.04}_{-0.05}$  &$1.39^{+0.10}_{-0.15}$   &$0.90^{+0.07}_{-0.11}$ &$0.65^{+0.05}_{-0.10}$  &$0.61^{+0.08}_{-0.06}$   &$0.56^{+0.07}_{-0.08}$   & $0.59^{+0.07}_{-0.03}$  &$0.54^{+0.04}_{-0.03}$\\
\\
$\tau$ & $4.16^{+0.28}_{-0.21}$  &$4.56^{+0.12}_{-0.14}$   &$4.30^{+0.27}_{-0.20}$ &$4.33^{+0.18}_{-0.21}$  &$5.23^{+0.30}_{-0.19}$   &$4.92^{+0.34}_{-0.39}$   & $4.29^{+0.39}_{-0.45}$  &$3.54^{+0.34}_{-0.46}$\\
\\
$\Gamma$ &$1.77^{+0.03}_{-0.03}$  &$1.74^{+0.02}_{-0.02}$   &$1.66^{+0.04}_{-0.05}$ &$1.73^{+0.02}_{-0.04}$  &$1.72^{+0.02}_{-0.04}$   &$1.70^{+0.03}_{-0.02}$   & $1.72^{+0.02}_{-0.02}$  &$1.66^{+0.03}_{-0.03}$\\
\\
$f_{\rm PL}$ &$0.84^{+0.01}_{-0.02}$  &$0.79^{+0.01}_{-0.02}$   &$0.81^{+0.03}_{-0.02}$ &$0.84^{+0.03}_{-0.03}$  &$ 0.88^{+0.03}_{-0.03}$  &$0.83^{+0.03}_{-0.04}$   & $0.88^{+0.03}_{-0.04}$  &$0.94^{+0.03}_{-0.03}$\\
\\
$\chi^2$/dof &1070/995                &2784/2551                &           2311/2120 &             1677/1643&              706/649    &            827/831    &             526/535  &            571/560  \\
\hline
\end{tabular}
\leftline{$\xi$'s are in the unit ergs cm s$^{-1}$. $\dagger$ pegged at the lowest value.}
\end{table*}


\subsubsection{Model--4 : RELXILL}

Reprocessed X-ray radiation is a feature often observed in AGN spectra. This reflection component typically consists of a reflection hump at $\sim 15-40$~keV and fluorescent iron lines. In Model 1 -- 3 we did not include a reprocessed X-ray radiation, hence, we add a relativistic reflection component {\tt RELXILL} \citep{Garcia2014,Dauser2014,Dauser2016} to our baseline model. 

In this model, the strength of reflection is measured from the relative reflection fraction ($R_{\rm refl}$), which is defined as the ratio between the observed Comptonized emission and the radiation reprocessed by the disc. {\tt RELXILL} assumes a broken power-law emission profile ($E(r) \approx r^{-q}$), where $r$ is the distance from the SMBH, $E(r)$ is the emissivity, and $q$ is the emissivity index. At a larger disc radii, in a non-relativistic domain, the emissivity profile has a form of  $E(r) \sim r^{-3}$. However, in the relativistic domain, the emissivity profile is steeper. The break radius ($R_{\rm br}$) separates the relativistic and non-relativistic domains. In our analysis, we fixed $q_2=3$ for emission at $r>R_{\rm br}$.

In our spectral analysis, we used {\tt RELXILL} along with the absorbed {\tt power-law} continuum. We only considered the reflection component from {\tt RELXILL} model by setting $R_{\rm refl}$ to a negative value. The model (hereafter Model--4) read in {\tt XSPEC} as,

{\tt TB $\times$ zxipcf1 $\times$ zxipcf2 $\times$ (zPL1 + RELXILL + bbody)}.

\begin{figure}
\centering
\includegraphics[angle=270,width=8.5cm]{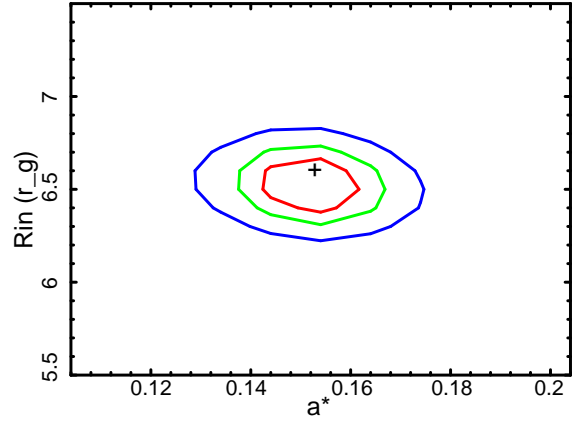}
\caption{2D contour plot between $R_{\rm in}$ and $a^*$ for Model--4 during the observation O2.}
\label{fig:rin-a}
\end{figure}

While fitting the data with this model, we tied the photon indices of the {\tt RELXILL} component to that of the {\tt power-law} model. Although, we started our analysis with one absorption component, we required two absorption component during O1, O2 and O3. The second absorption component improved the fit with $\Delta \chi^2 =$~82, 68 and 59 for 3 dof, during O1, O2 and O3 respectively. Throughout the observations, we obtained a fairly unchanged value of the iron abundances with $A_{\rm Fe} \sim 3.7-4.2$ $A_{\sun}$. Disc ionization was also constant during our observation period with $\xi \sim 10^{1.9-2.2}$ erg~cm~s$^{-1}$. The emissivity profile was quite stable with $q_2 \sim 4-5$, although $R_{\rm br}$ was found to change. This parameter reached its maximum during the observation~O1 ($R_{\rm br} = 42^{+3}_{-2}$ $R_g$). In the later observations, it decreased and varied in the range of $R_{\rm br} \sim 16-26$ $R_g$. The inner edge of the disc varied in the range of $R_{\rm in} \sim 4-7$ $R_g$. The best-fit inclination angle of the AGN was obtained in the range of $10.3^{+4.9}_{-6.1}$\textdegree -- $17.7^{+2.9}_{-5.2}$\textdegree. The spin of the BH in \source~ was observed to be low, with the best-fitted spin parameters found to be in the range of $0.15^{+0.03}_{-0.03} - 0.21^{+0.04}_{-0.02}$, which is consistent with what we found from the {\tt OPTXAGNF} model. In all the observations, the reflection component was found to be relatively weak with reflection fraction varied in the range of, $R_{\rm refl} \sim 0.10-0.18$. The {\tt RELXILL} model fitted spectral analysis results are given in Table~\ref{tab:rel}. We show the unfolded spectrum fitted with Model--4 for observation O2 in Figure~\ref{fig:uf-spec}. In Fig.~\ref{fig:rin-a}, we show the contour plot of $R_{\rm in}$ and $a^*$ for the observation O2.

\begin{table*}
\caption{Best-fit parameters obtained from the spectral fitting of the source spectra with Model--4 ({\tt RELXILL}).}
\label{tab:rel}
\hspace*{-0.5in}
\begin{tabular}{lccccccccc}
\hline
&X1 &O1   &O2    & O3  &O4     &O5  &O6 &O7                        \\
\hline
$N_{\rm H,1}$ ($10^{21}$ \pcm) & $3.51^{+0.03}_{-0.04}$  & $0.77^{+0.06}_{-0.11}$  &$0.95^{+0.11}_{-0.14}$ &$1.18^{+0.12}_{-0.10}$  &$1.28^{+0.08}_{-0.10}$   &$1.16^{+0.07}_{-0.06}$   &$1.28^{+0.10}_{-0.08}$   &$1.31^{+0.12}_{-0.18}$\\
\\
log${\rm \xi_{\rm 1}}$ &$-3^{\dagger}$          & $1.63^{+0.05}_{-0.04}$  &$1.85^{+0.05}_{-0.04}$ &$1.31^{+0.07}_{-0.08}$  &$1.05^{+0.05}_{-0.08}$   &$0.21^{+0.05}_{-0.06}$   &$0.19^{+0.06}_{-0.03}$   &$0.24^{+0.04}_{-0.03}$\\
\\
Cov Frac2 &$0.17^{+0.05}_{-0.06}$  & $0.44^{+0.04}_{-0.06}$  &$0.36^{+0.05}_{-0.08}$ &$0.33^{+0.06}_{-0.12}$  &$<0.14$                  &$<0.18$   &$<0.16$& $<0.13$ \\
\\
$N_{\rm H,2}$ ($10^{21}$ \pcm) & --                     & $4.36^{+0.34}_{-0.47}$  &$4.58^{+0.37}_{-0.42}$ &$0.95^{+0.19}_{-0.26}$  & --                      & -- & -- & --\\
\\
log${\rm \xi_{\rm 2}}$ &--                      & $4.58^{+0.41}_{-0.26}$  &$3.10^{+0.19}_{-0.24}$ &$2.96^{+0.26}_{-0.29}$  & -- & -- & --& -- \\
\\
Cov Frac2 &--                      &$>0.74$                  &$0.52^{+0.18}_{-0.23}$ &$0.53^{+0.31}_{-0.25}$  & -- & --&-- & -- \\
\\
$\Gamma$ &$   1.77^{+0.03}_{-0.04}$   &$   1.76^{+0.05}_{-0.03}$  &$   1.68^{+0.04}_{-0.02}$  &$   1.73^{+0.03}_{-0.03}$  &$   1.65^{+0.04}_{-0.03}$   &$   1.67^{+0.05}_{-0.03}$ &$   1.71^{+0.02}_{-0.03}$ &$   1.68^{+0.03}_{-0.04}$ \\
\\
$N_{\rm PL}$ (10$^{-3}$ \phc)&$   1.48^{+0.10}_{-0.14}$   &$  22.35^{+1.02}_{-0.93}$  &$   6.15^{+0.45}_{-0.68}$  &$   2.54^{+0.32}_{-0.47}$  &$   4.41^{+0.41}_{-0.57}$   &$   2.82^{+0.26}_{-0.32}$ &$   2.72^{+0.15}_{-0.22}$ &$   2.55^{+0.21}_{-0.28}$ \\
\\
$A_{\rm Fe}$ ($A_{\odot}$)&$ 3.68^{+0.22}_{-0.28}$ &$ 3.89^{+0.26}_{-0.35}$ &$ 4.06^{+0.32}_{-0.22}$ &$ 4.15^{+0.34}_{-0.38}$ &$ 3.77^{+0.27}_{-0.38}$ &$ 4.12^{+0.27}_{-0.37}$ &$ 3.95^{+0.38}_{-0.42}$ &$ 4.18^{+0.27}_{-0.39}$ \\
\\
log($\xi$)&$ 1.96^{+0.03}_{-0.02}$ &$2.09^{+0.05}_{-0.04}$ &$ 2.18^{+0.03}_{-0.02}$ &$ 2.11^{+0.02}_{-0.02}$ &$ 2.07^{+0.02}_{-0.03}$ &$2.18^{+0.04}_{-0.03}$ &$ 2.13^{+0.02}_{-0.03}$ &$ 1.98^{+0.02}_{-0.02}$ \\
\\
$\theta_{\rm incl}$ (degree)&$   13.8^{+2.7}_{-4.1}$ &$11.8^{+4.1}_{-5.5}$  &$ 16.2^{+2.8}_{-4.7}$  &$ 14.3^{+4.4 }_{-3.9 }$ &$ 10.3^{+4.9 }_{-6.1}$ &$ 13.2^{+2.5}_{4.7}$ &$ 17.7^{+2.9}_{-5.2}$ &$ 12.1^{+4.5}_{-5.8}$ \\
\\
$R_{\rm refl}$ &$  0.11^{+0.02}_{-0.02}$   &$ 0.16^{+0.03}_{-0.04}$  &$ 0.15^{+0.03}_{-0.02}$ &$ 0.13^{+0.02}_{-0.02}$ &$ 0.18^{+0.02}_{-0.03}$ &$0.16^{+0.03}_{-0.02}$ &$ 0.10^{+0.03}_{-0.02}$ &$ 0.12^{+0.03}_{-0.02}$ \\
\\
$q_2$&$ 3.66^{+0.17}_{-0.41}$ &$ 4.69^{+0.68}_{-0.35}$&$ 4.41^{+0.25}_{-0.37}$ &$ 5.35^{+0.45}_{-0.29}$ &$ 5.31^{+0.29}_{-0.41}$ &$4.81^{+0.42}_{-0.63}$ &$  4.56^{+0.72}_{-0.89}$ &$  4.94^{+0.60}_{-0.76}$ \\
\\
$R_{\rm br}$ ($R_g$)&$ 12^{+3 }_{-2 }$ &$ 42^{+3}_{-2}$  &$ 26^{+2}_{-3}$  &$ 17^{+3}_{-1}$ &$ 18^{+2}_{-3}$ &$ 21^{+2}_{-3}$ &$ 21^{+2}_{-1}$ &$ 16^{+2}_{-3}$ \\
\\
$a^*$&$0.15^{+0.02}_{-0.03}$   &$0.16^{+0.03}_{-0.03}$  &$0.15^{+0.03}_{-0.03}$  &$0.17^{+0.02}_{-0.03}$  &$0.21^{+0.04}_{-0.02}$   &$0.15^{+0.03}_{-0.03}$ &$0.16^{+0.02}_{-0.03}$ &$0.20^{+0.02}_{-0.04}$ \\
\\
$R_{\rm in}$ ($R_g$)&$   4.71^{+0.65}_{-0.88}$ &$ 5.78^{+0.43}_{-0.77}$ &$ 6.61^{+0.33}_{-0.37}$  &$   4.76^{+0.28}_{-0.48}$ &$ 6.73^{+0.59}_{-0.73}$  &$   4.06^{+0.32}_{-0.62}$ &$ 4.88^{+0.74}_{-0.91}$ &$   6.92^{+0.65}_{-0.85}$ \\
\\
$N_{\rm rel}$ (10$^{-5}$ \phc)&$   0.62^{+0.08}_{-0.10}$   &$ 7.11^{+0.42}_{-0.62}$  &$ 0.49^{+0.06}_{-0.07}$ &$  1.18^{+0.03}_{-0.04}$  &$ 1.87^{+0.14}_{-0.23}$  &$   0.67^{+0.07}_{-0.10}$ &$ 1.10^{+0.12}_{-0.06}$ &$   0.56^{+0.04}_{-0.07}$ \\
\\
$\chi^2$/dof&  983/992 & 2845/2550 & 2275/2119 &1648/1640 & 685/646 & 829/828 & 541/532   & 533/557 \\ 
\hline
\end{tabular}
\leftline{$\xi$'s are in the unit ergs cm s$^{-1}$. $\dagger$ pegged at the lowest value.}
\end{table*}

\section{Discussion}

We studied NGC~1566 during and after the 2018 outburst event using data from \xmm, \swift~ and  \nustar~ in the $0.5-70$~keV energy band. From a detailed spectral and timing analysis, we explored the nuclear properties of the AGN.

\subsection{Black Hole Properties}

\source~ hosts a supermassive black hole of mass $M_{\rm BH} \approx 8.3 \times 10^6$ $M_{\odot}$ \citep{Woo2002}. We kept the mass of the BH frozen during our spectral analysis with Model--3. Fitting the spectra with Model--3 and Model--4, we estimated the spin parameter ($a^*$) to be in the range $0.18^{+0.01}_{-0.02}-0.21^{+0.03}_{-0.02}$ and $0.15^{+0.2}_{-0.3}-0.21^{+0.04}_{-0.02}$, respectively. Both models favour a low spinning BH with the spin parameter $a^* \sim 0.2$, which is consistent with the findings of \citet{Parker2019}.

The inclination angle was a free parameter in Model--4, and it was estimated to be in the range of $10.3^{+4.9}_{-6.1}$\textdegree -- $17.7^{+2.9}_{-5.2}$\textdegree. \citet{Parker2019} also found a consistent inclination angle ($\theta_{\rm incl} <$ 11\textdegree). 

\begin{table*}
\caption{Luminosities of \source\ in the observations analyzed here.}
\label{tab:lum}
\begin{tabular}{lccccccccc}
\hline
ID & Day & $L_{\rm nuc}$ & $L_{\rm soft}$ & $L_{\rm 0.1-100}$ & $L_{\rm bol}$ & $\lambda_{\rm Edd}$ \\
 & & ($10^{42}$ \eps) & ($10^{41}$ \eps) & ($10^{42}$ \eps) & ($10^{43}$ \eps) & \\
\hline
X1&  57331  &$   0.34 \pm0.01    $&$  0.89 \pm0.06    $&$   0.43 \pm0.01    $&$    0.09 \pm0.01  $&$  0.003\pm0.001 $\\
O1&  58295  &$  13.42 \pm0.08    $&$  4.89 \pm0.33    $&$  13.91 \pm0.08    $&$   7.11 \pm0.04  $&$   0.066\pm0.001 $\\
O2&  58395  &$   4.02 \pm0.05    $&$  2.45 \pm0.14    $&$   4.26 \pm0.05    $&$    1.84 \pm0.03  $&$  0.017\pm0.001 $\\
O3&  58639  &$   2.36 \pm0.01    $&$  1.16 \pm0.25    $&$   2.48 \pm0.03    $&$    1.19 \pm0.02  $&$  0.011\pm0.001 $\\
O4&  58703  &$   4.01 \pm0.05    $&$  1.56 \pm0.22    $&$   4.17 \pm0.05    $&$    2.04 \pm0.03  $&$  0.019\pm0.001 $\\
O5&  58706  &$   2.16 \pm0.06    $&$  1.64 \pm0.21    $&$   2.32 \pm0.06    $&$    1.33 \pm0.05  $&$  0.012\pm0.001 $\\
O6&  58713  &$   1.96 \pm0.12    $&$  1.54 \pm0.29    $&$   2.11 \pm0.12    $&$    1.29 \pm0.07  $&$  0.012\pm0.001 $\\
O7&  58716  &$   2.40 \pm0.03    $&$  1.59 \pm0.31    $&$   2.58 \pm0.04    $&$    1.31 \pm0.03  $&$  0.012\pm0.001 $\\
\hline
\end{tabular}
\leftline{$L_{\rm nuc}$ and $L_{\rm soft}$ are calculated for the primary power-law and soft excess components, respectively.}
\leftline{Eddington ratio, $\lambda_{\rm Edd}$ is calculated using $L_{\rm bol}/L_{\rm Edd}$ for a BH of mass $8.3 \times 10^6$ \M.}
\end{table*}

\subsection{Corona Properties}

The X-ray emitting corona is generally located very close to the central BH \citep{Fabian2015}. This corona is characterized by the photon index ($\Gamma$), temperature ($kT_{\rm e}$) and optical depth ($\tau$) of the Comptonizing plasma. While using a simple power-law model gives us information only about the photon index, the {\tt NTHCOMP} model can provide us with information on the electron temperature of the Compton cloud ($kT_{\rm e}$), while the optical depth ($\tau$) is calculated from Equation~\ref{eqn:tau}. We found that the photon index varied within a narrow range of $\Gamma \sim 1.7-1.8$, and can be considered as constant within the uncertainties. To constrain the photon index with more accuracy, we fitted the source spectra from \nustar~ observations in $3-70$~keV and $10-70$~keV energy ranges to approximate only the power-law part. We found similar results as from the simultaneous fitting of the \xmm~ and \nustar~ data in $0.5-70$~keV range. From the spectral analysis with Model--3, we estimated the Compton cloud radius to be $12-43~R_{\rm g}$.

We calculated the intrinsic luminosity ($L_{\rm 0.1-100}$) of the AGN from Model--1. The intrinsic luminosity was observed to be low during the pre-outburst observation, X1, with $L_{\rm 0.1-100} \sim (4.3 \pm 0.1) \times 10^{41}$ \eps. During O1,  was observed to be maximum with $L_{\rm 0.1-100} \sim (1.39 \pm 0.01) \times 10^{44}$ \eps. Later, the intrinsic luminosity decreased and varied in the range of $2.1-4.3 \times 10^{43}$ \eps. We also computed the luminosity for the primary emission ($L_{\rm nuc}$) and soft excess ($L_{\rm soft}$) from the individual components while analyzing the spectra with Model--1. We calculated the bolometric luminosity ($L_{\rm bol}$) using the 2--10\,keV bolometric correction, $\kappa_{\rm bol, 2-10~keV} = 20$ \citep{Vasedevan2009}. The Eddington ratio ($\lambda_{\rm Edd}=L_{\rm bol}/L_{\rm Edd}$), assuming a BH of mass of $8.3 \times 10^6$ \M \citep{Woo2002}, was estimated to be $\lambda_{\rm Edd} \sim 0.003-0.066$ in different epoch which is consistent with other nearby Seyfert-1 galaxies \citep{Wu2004,Sikora2007,Koss2017}. 

In the pre-outburst observation in November 2015 (X1 : see Table~\ref{tab:lum}), we obtained a bolometric luminosity of $L_{\rm bol}=(0.9\pm 0.1) \times 10^{42}$ \eps, with corona size $R_{\rm cor} = 12\pm 3~R_{\rm g}$ and hot electron plasma temperature $kT_{\rm e}=102\pm 5$~keV. In the observation during the outburst in June 2018 (O1), the luminosity of the AGN increased by a factor of about $\sim$25, compared to the November 2015 observation (X1). During this observation, the corona was large ($R_{\rm cor} = 43\pm 3~R_{\rm g}$) with hot electron plasma temperature $kT_{\rm e} = 61\pm7$~keV and the observed spectrum was harder. As the outburst progressed, the bolometric luminosity and the corona size decreased. As $R_{\rm cor}$ decreased, the electron plasma temperature increased. Overall, $kT_{\rm e}$ varied in a range of $\sim 61-106$~keV during the observations. In general, the plasma temperature is observed in a wide range, with a median at $kT_{\rm e} \sim 105 \pm 18$~keV \citep{Ricci2018}. Thus, the plasma temperature is consistent with other AGNs. During these observations, the optical depth of the Compton cloud varied within $\sim 1.2-1.7$. Interestingly, the photon index ($\Gamma$) was almost constant, although some of the properties of the corona evolved with time. This appears to imply that both the optical depth and the hot electron temperature changed in such a way that the spectral shape remained the same. 

\begin{figure*}
\centering
\includegraphics[width=16cm, angle=0]{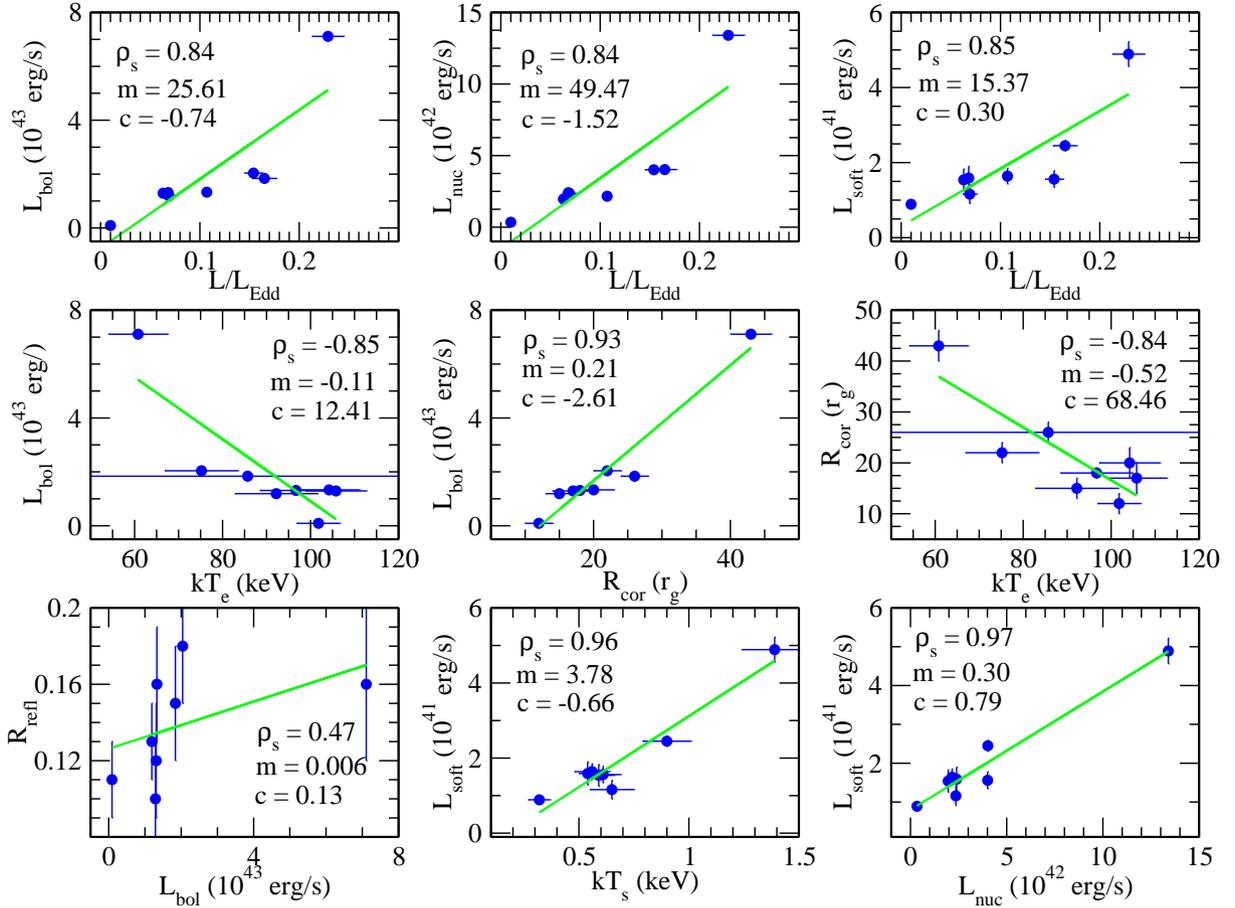}
\caption{Correlation between different spectral parameters. In each panel, the corresponding Pearson correlation co-efficient ($\rho_{s}$) is quoted. The solid green line in each panel represent the linear fit, y=mx+c. Corresponding fitted values of the slope (m) and intercept (c) are also mentioned in each panels.}
\label{fig:cor}
\end{figure*}

We found several correlations and anti-correlations between the spectral parameters and show them in Fig.~\ref{fig:cor}. We fitted the data points with linear regression method using y = mx + c. The fitted value of the slope (m) and intercept (c) are mentioned in each panel of Fig.~\ref{fig:cor}. We found that the nuclear luminosity ($L_{\rm nuc}$) and the soft excess luminosity ($L_{\rm soft}$) are strongly correlated with the Eddington ratio with the Pearson correlation indices of 0.84 and 0.85, respectively. We also found that the bolometric luminosity ($L_{\rm bol}$) and the Compton cloud temperature ($kT_{\rm e}$) are anti-correlated ($\rho_{\rm s}=-0.85$). The electron temperature is found to be anti-correlated with the Eddington ratio ($\rho_{\rm s}=-0.85$), while the size of the Compton corona and the luminosity are positively correlated ($\rho_{\rm s}=0.93$). We also observed that the electron temperature is anti-correlated with the size of the Compton cloud ($\rho_{\rm s}=-0.84$). The above correlations can be explained thinking that, as the accretion rate increased, the energy radiation increased, thereby, increasing the luminosity. An increase in the mass accretion rate makes the cooling more efficient, leading to a decrease in the electron temperature of the Comptonizing region \citep{HM1991,Done2007}.

\subsection{Reflection}
The hard X-ray photons from the corona are reflected from cold material in the accretion disc, BLR and torus, producing a reflection hump and a Fe-K emission line \citep{George1991,Matt1991}. When fitted with a simple power-law model, \source~ showed the presence of Fe K$\alpha$ emission line along with a weak reflection hump at $\sim 15-40$~keV energy range (see Fig~\ref{fig:uf-spec}). Thus, we fitted the spectra with the relativistic reflection model {\tt RELXILL} to probe the reflection component.

\citet{Parker2019} analyzed O1 observation with {\tt RELXILL} and {\tt XILLVER} models in their spectral analysis. They found $\xi = 10^{2.4\pm0.1}$ \ecs, $A_{\rm Fe} = 3\pm0.2 A_{\odot}$ and $R_{\rm refl} = 0.091\pm0.005$. In the present work, we obtained $\xi= 10^{2.09\pm0.05}$ \ecs, $A_{\rm Fe} = 3.89\pm0.35$, and $R_{\rm refl} = 0.16\pm0.04$ for O1. The marginal difference between our results and those of \citet{Parker2019} could be ascribed to the different spectral models used. We observed fairly constant ionization ($\xi \sim 10^{1.9-2.2}$ erg cm s$^{-1}$) and iron abundances ($A_{\rm Fe} \sim 4-5$ $A_{\sun}$) across the observations. This is expected within our short period of observation. In all the spectra, we found a very weak reflection with reflection fraction, $R_{\rm refl} < 0.2$. We found a weak correlation between the reflection fraction ($R_{\rm refl}$) and the luminosity with the Pearson correlation coefficient of 0.47. In general, reflection becomes strong with increase in the luminosity \citep{Z99}. The low inclination angle of the source also results in a weak reflection \citep{Ricci2011,Chatterjee2018}. Therefore, the observed weak correlation between the reflection fraction and luminosity in \source\ could be due to the low inclination angle of the source.

During the observations, a strong Fe K$\alpha$ emission line with equivalent width $EW > 100$~eV was detected, despite a weak reflection component, except for O5. This could be explained by high iron abundances in the reflector. From the spectral analysis with model--4, we found the inner edge of the disc extends up to $\sim 5~R_{\rm g}$. If iron originates from the inner disc, a broad iron line is expected. However, we did not observe a broad iron line. Either the broad line was absent or it was blurred beyond detection. However, during our observation, a narrow iron line was detected, which given its width, originates in the material further out than the accretion disc. Hence, from the full-width at half maximum (FWHM) of the line, we tried to constraint the Fe K$\alpha$ line emitting region. During our observation period, the FWHM of Fe K$\alpha$ line emission was found to be $<8700$ km s$^{-1}$, which corresponds to the region $>1200$~$R_{\rm g}$ from the BH. This corresponds to the distance at which we expect to find the BLR \citep{Kaspi2000}. Thus, the BLR is the most probable Fe K$\alpha$ line emitting region in NGC~1566.

\subsection{Soft Excess}

The origin of the soft excess in AGNs is still very debated, and several models have been proposed to explain it. Relativistic blurred ionized reflection from the accretion disc has been put forward as a likely explanation for the soft excess in many sources \citep{Fabian2002,Ross2005,Walton2013,Garcia2019,Ghosh2020}. An alternative scenario considers Comptonization by a optically thick, cold corona \citep{Done2012}. In this model, the comptonizing region is located above the inner accretion disc as a thin layer. Heating of circumnuclear region by bulk motion Comptonization in AGN with high Eddington ratio is also considered to be the reason for the soft excess \citep{Kaufman2017}. Recently, \cite{Nandi2021} argued from long-term observations and Monte-Carlo simulations (see \cite{Chatterjee2018} and references therein) that the thermal Comptonization of photons which have suffered fewer scatterings could explain the origin of soft excess in Ark 120.     

In our analysis, we used a {\tt bbody} component to take into account the soft excess. The blackbody temperature ($kT_{\rm bb}$) was roughly constant with $kT_{\rm bb} \sim 110$~eV during our observations. This is consistent with the observation of other nearby AGNs \citep{Gierlinski2004,Winter2009,Ricci2017,Garcia2019}. A good-fit with the {\tt OPTXAGNF} model favoured soft Comptonization by an optically thick corona as the origin of the soft excess. During observation~O1, the temperature of the optically-thick corona was observed to be $\sim 1.4$~keV. Later, the optically thick corona cooled with decreasing luminosity. 

We tried to estimate delay and correlation between the soft excess and X-ray continuum light curves (see Section~\ref{sec:correlation}). In the {\tt OPTXAGNF} scenario, the total energetic depends on the mass accretion rate and is divided between the soft-Comptonization (soft-excess) and hard-Comptonization (nuclear or primary emission). We found a strong correlation between the soft excess luminosity ($L_{\rm soft}$) and nuclear luminosity ($L_{\rm nuc}$). The soft-excess luminosity and the Eddington ratio ($L/L_{\rm Edd}$) were found to be correlated ($\rho_{\rm s}=0.78$). This indicated that the soft excess strongly depended on the accretion rate, supporting the soft-Comptonization as the origin of the soft excess. We also found a delay of $\sim 10$ minutes between the soft excess and primary emission, indicating the origin of the soft excess was beyond the corona, possibly the accretion disc. Higher variability was also observed in the soft excess (see Section~\ref{sec:correlation}) during observations X1, O1, and O3. This could indicate a higher stochasticity in the origin of the soft-excess. It should be noted that, theoretically, infinite scattering produces blackbody which has the lowest variability. As \citet{Nandi2021} suggested, fewer scattering which could generate higher variability than the large number of scatterings, possibly lead to the origin of soft-excess. Overall, the soft excess in \source~ has a complex origin, including reflection from the accretion disc and soft-Comptonization.

\subsection{Changing-Look Event and its Evolution}

The physical drivers of CL events are still highly debated, and could change from source to source. Tidal disruption events (TDEs), changes in obscuration, and variations in the mass accretion rate could all be possible explanations for CL events. 

In the pre-outburst observation, X1, the absorber was not strongly ionized ($\xi_1 <0.001$), and had a column density of $N_{\rm H,1} = (3.53 \pm 0.06) \times 10^{21}$ \pcm. Two ionizing absorbers were detected during the observation O1, O2, and O3 which could be associated with an outflow. \citet{Parker2019} also observed two ionized absorbers during O1, and suggested that they could be associated with an outflow. The highest ionization (for both absorbers) was observed in O1, which corresponded to the epoch in which the observed luminosity was the highest. During the 2018 outburst, when the X-ray intensity increased, the emitted radiation may have caused the sublimation of the dust along the line of sight, thereby decreasing the hydrogen column density \citep{Parker2019}. As the X-ray intensity decreased after the main outburst, the dust could have condensed, leading to an increase in the column density. Eventually, during the August 2019 observations, the dust formation was stable as the hydrogen column density was approximately constant ($N_{\rm H} \sim 1.3 \times 10^{21}$ \pcm). Generally, the dust clouds can recover in the timescale of several years \citep{Kishimoto2013,Oknyansky2017}. In this case, we observed that the dust clouds already recovered with an increase in $N_{\rm H}$ from $\sim 6 \times 10^{20}$ \pcm~ to $1.3\times 10^{21}$ \pcm~ in $\sim$14 months time. If this is correct, we would expect the $N_{\rm H}$ to reach at its pre-outburst value of $N_{\rm H} \sim 3.5 \times 10^{21}$ \pcm~ in next few months.

The strong correlation between the accretion rate and the bolometric luminosity suggests that the accretion rate is responsible for the CL events in \source~ during the 2018 outburst. \citet{Parker2019} also discussed several possibilities for the CL event in \source~ and concluded that the disc-instability is the most likely reason for the outburst. The instability at the outer disc could propagate through the disc and cause the outburst. \citet{Noda2018} explained the flux drop and changing-look phenomena in Mrk~1018 with the disc-instability model where the time-scale for the changing-look event was $\sim 8$~years. The time-scale for changing-look event for NGC~1566 was $\sim 10$~months as the flux started to increase from September 2017 \citep{Dai18,Cutri18}. The time-scales are similar if we consider the mass of the BH. As the mass of Mrk~1018 is $M_{\rm 1018} \sim 10^{7.84}~M_{\odot}$ \citep{Ezhikode2017}, the expected time-scale for NGC~1566 is $\sim 8~{\rm years}/10 \sim 10$~months. The observed `q'-shaped in the HID during the main outburst (F1), also suggest of disc instability as seen in the case of the outbursting black holes. The `q'-shaped HID is very common for the outbursting black holes \citep{RM06} where disc instability is believed to lead the outburst. \citet{Noda2018} also suggested that the soft-excess would decrease much more compared to the hard X-ray with decreasing $L/L_{\rm Edd}$. NGC~1566 also showed the strongest soft-excess emission during the highest $L/L_{\rm Edd}$ (O1), while the soft-excess emission dropped as $L/L_{\rm Edd}$ decreased. In the pre-outburst quiescent state, no soft-excess was observed in NGC~1566. As the soft-excess can produce most of the ionizing photons necessary to create broad optical lines \citep{Noda2018}, the broad line appeared during O1 (when the soft-excess was strong), leading to the changing-look event.

\begin{figure}
\centering
\includegraphics[angle=0,width=8.5cm]{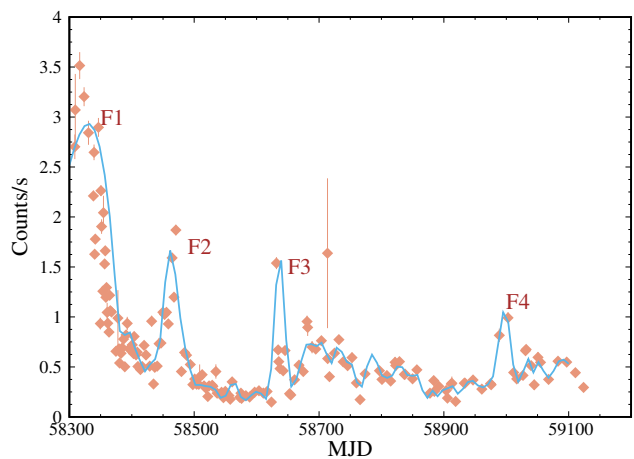}
\caption{Lightcurve of NGC 1566 between June 2018 and August 2020 with smooth lines showing near-periodic variation of count rate.}
\label{fig:lc-smooth}
\end{figure}

A TDE might be another possible explanation for the 2018 outburst of NGC\,1566. A TDE could supply the accreting matter to the central SMBH, which would lead to an increase in luminosity. Several recurrent outbursts (F2, F3 \& F4) and nearly periodic X-ray variations were observed after the main outburst (F1) as seen in the upper panel of Fig.~\ref{fig:xrt-lc} and Fig~\ref{fig:lc-smooth}. In TDEs, some amount of matter could be left out, and cause recurrent outbursts \citep{Komossa2017}. In the case of a classic TDE, after a star is tidally disrupted by the SMBH, a decay profile of the luminosity with $t^{-5/3}$ is expected \citep{Rees1988,komossa2015,Komossa2017}, which is not observed in this case. During all observations, the source showed a relatively hard spectrum, with $\Gamma \sim 1.6-1.7$. This is clear contrast with classic TDEs, which typically show much softer spectra ($\Gamma \geq 3$) \citep{Komossa2017}. During the June 2018 outburst (F1), the X-ray luminosity changed by about $\sim 25$~times compared to the low state, which is low in comparison to other candidate TDEs. For example, 1ES~1927+654 and RX~J1242-1119 showed a change in luminosity by over 4 orders of magnitude \citep{Ricci2020} and 1500 times \citep{Komossa2004}, respectively. In general, no iron emission line is observed in the X-ray band during the TDE \citep{Saxton2020}, whereas the case of NGC\,1566 a strong Fe K$\alpha$ line was observed. Considering all this, we deem unlikely that the June 2018 outburst of \source~ was triggered by a TDE.

An alternative explanation could be that a star is tidally disrupted by a merging SMBH binary at the center of NGC~1566. According to \cite{Hayasaki2016}, after the stellar tidal disruption, the debris chaotically moves in the binary potential and is well-stretched. The debris orbital energy is then dissipated by the shock due to the self-crossing, leading to the formation of an accretion disc around each black hole after several mass exchanges through the Lagrange (L1) point of the binary system. If the orbital period of an unequal-mass binary is short enough ($<1000$~days) to lose the orbital energy by gravitational wave (GW) emission, the secondary, less massive SMBH orbits around the center of mass at highly relativistic speed, while the primary SMBH hardly moves. Therefore, the electromagnetic emission from the accretion disc around the secondary black hole would be enhanced periodically by relativistic Doppler boosting. This scenario could explain the recurrent outburst observed in case of NGC~1566 (see Fig~\ref{fig:lc-smooth}). From this figure, the orbital period is estimated to be $P_{\rm orb}\sim 160$ days (between F1 \& F2). Taking into account the SMBH mass ($M_{\rm BH}=8.3\times10^6 ~{\rm M_{\odot}}$), we obtain $a\sim 710~r_{\rm s} \approx 5.6 \times 10^{-4}$~pc as a binary semi-major axis, where $r_{\rm s}=2GM_{\rm BH}/c^2$ is the Schwarzschild radius. The merging timescale of two SMBH with mass ratio $q$ due to GW emission is given by \citep{Peters1964},

\begin{equation}
t_{\rm gw} = \frac{5}{8} \frac{(1-q)^2}{q} \frac{r_s}{c} \Big( \frac{a}{r_s} \Big)\sim 5.0\times10^6 yr.
\end{equation}

This suggests that more than 100 TDEs could occur before the SMBH merger, considering the TDE rate for a single SMBH ($10^{-4}$ to $10^{-5}$ $yr^{-1}$ per galaxy, \citealp{Stone2020}). However, the event rate can be enhanced up to 0.1 $yr^{-1}$ per galaxy due to chaotic orbital evolution and Kozai-Lidov effect in the case of SMBH binaries \citep{Chen2009,Li2015}. Moreover, if stars are supplied by accretion from a circumbinary disc \citep[eg,][]{Hayasaki2007,Amaro2013}, then the TDE rate could be higher up to $\sim0.2~ yr^{-1}$ if the mass supply rate is at the Eddington limit \citep{Wolf2021}. Therefore, the detection of similar, periodic burst events in the next few years to tens of years would support this interpretation.

\section{Summary}

We analyzed the X-ray emission of the changing-look AGN \source\ between 2015 and 2019. Our key findings are the following.

\begin{enumerate}
    \item \source\ showed a giant outburst in June 2018 when the X-ray luminosity increased by $\sim25-30$ times compared to that during the low state. After the main outburst, several recurrent outbursts were also observed.
    \item \source~ hosts a low-spinning BH with the spin parameter, $a^* \sim 0.2$.
    \item The inclination angle is estimated to be in the range of $i \sim 10$\textdegree--$21$\textdegree.
    \item The variation of the accretion rate is responsible for the evolution of the Compton corona and X-ray luminosity.
    \item A rise in the accretion rate is responsible for the change of luminosity. The HID or `q'-diagram links the CL event of \source~ with the outbursting black holes. 
    \item A strong soft excess was observed when the luminosity of \source~ was high. The origin of the soft excess is observed to be complex. 
    \item We rule out the possibility that the event was triggered by a classical TDE, where a star is tidally disrupted by the SMBH. 
    \item We propose a possible scenario where the central core is a merging binary SMBH. This scenario could explain the recurrent outburst. 
\end{enumerate}

\section*{Data Availability}

We have used archival data for our analysis in this manuscript. All the models and software used in this manuscript are publicly available. Appropriate links are given in the text.

\section*{Acknowledgements}
We acknowledge the anonymous reviewer for the helpful comments and suggestions which improved the paper. Work at Physical Research Laboratory, Ahmedabad, is funded by the Department of Space, Government of India. PN acknowledges CSIR fellowship for this work. A. C. and K.H. has been supported by the Basic Science Research Program through the National Research Foundation of Korea (NRF) funded by the Ministry of Education (2016R1A5A1013277 (A.C. and K.H) and 2020R1A2C1007219 (K.H.)). K.H. acknowledges the Yukawa Institute for Theoretical Physics (YITP) at Kyoto University. Discussions during the YITP workshop YITP-T-19-07 on International Molecule-type Workshop ``Tidal Disruption Events: General Relativistic Transients" were useful for this work. This research has made use of data and/or software provided by the High Energy Astrophysics Science Archive Research Center (HEASARC), which is a service of the Astrophysics Science Division at NASA/GSFC and the High Energy Astrophysics Division of the Smithsonian Astrophysical Observatory. This research has made use of the {\it NuSTAR} Data Analysis Software ({\tt NuSTARDAS}) jointly developed by the ASI Space Science Data Center (SSDC, Italy) and the California Institute of Technology (Caltech, USA). This research has made use of the SIMBAD database, operated at CDS, Strasbourg, France.



\bibliographystyle{mnras}
\bibliography{ref-ngc1566} 




\appendix

\begin{table*}
\caption{\swift/XRT observation log}
\label{tab:analysis}
\begin{tabular}{|ccc|ccc|ccc|}
\hline
Obs ID & Date & Exp (s) & Obs ID & Date & Exp (s) & Obs ID & Date & Exp (s) \\
\hline
00035880002& 	2007-12-12& 2151 &00035880060& 	2018-10-19& 677	&00035880122&	2019-06-02& 709    \\ 
00035880003&	2008-01-02& 3230 &00035880061&	2018-10-22& 967	&00035880123&	2019-06-03& 694    \\
00045604001&	2011-08-25& 378	 &00035880062&	2018-10-25& 934	&00035880125&	2019-06-04& 519    \\
00045604002&	2011-11-03& 1135 &00035880063&	2018-10-28& 1135&00035880126&	2019-06-05& 909    \\
00045604003&	2012-01-06& 142	 &00035880064&	2018-10-31& 937	&00035880127&	2019-06-06& 754    \\
00045604004&	2012-09-15& 1210 &00035880065&	2018-11-06& 957	&00035880128&	2019-06-07& 659    \\
00045604005&	2012-09-29& 2773 &00035880066&	2018-11-09& 1017&00035880129&	2019-06-08& 674    \\
00045604006&	2012-09-29& 4518 &00035880067&	2018-11-12& 992	&00035880130&	2019-06-09& 674    \\
00045604007&	2012-10-09& 4526 &00035880068&	2018-11-15& 917	&00035880131&	2019-06-10& 814    \\
00045604008&	2012-10-10& 1138 &00035880069&	2018-11-18& 900	&00035880132&	2019-06-11& 764    \\
00035880004& 	2018-06-24& 1352 &00035880070&	2018-11-21& 822	&00035880133&	2019-06-13& 369    \\
00035880005&	2018-07-10& 95	 &00035880071&	2018-11-24& 992	&00035880134&	2019-06-14& 974    \\
00035880006&	2018-07-10& 1030 &00035880072&	2018-11-27& 1113&00035880136&	2019-06-16& 1048   \\
00035880007&	2018-07-17& 254	 &00035880073&	2018-11-30& 1119&00035880137&	2019-06-19& 1053   \\
00035880008&	2018-07-17& 999	 &00035880074&	2018-12-03& 909	&00035880138&	2019-06-20& 920    \\
00035880009&	2018-07-24& 239	 &00035880075&	2018-12-06& 1024&00035880139&	2019-06-26& 543    \\
00035880010&	2018-07-24& 1059 &00035880076&	2018-12-12& 1084&00035880140&	2019-07-03& 892    \\
00035880013&	2018-07-31& 1073 &00035880077&	2018-12-15& 954	&00035880141&	2019-07-10& 915    \\
00035880014&	2018-08-08& 975	 &00035880078&	2018-12-18& 999	&00035880142&	2019-07-17& 1103   \\
00035880015&	2018-08-09& 1205 &00035880079&	2018-12-21& 994	&00035880143&	2019-07-24& 878    \\
00035880016& 	2018-08-10& 969	 &00035880080&	2018-12-24& 909	&00035880144&	2019-07-31& 942    \\
00035880017&	2018-08-11& 969	 &00035880081&	2018-12-27& 584	&00088910001&	2019-08-08& 1924   \\
00035880019&	2018-08-13& 1009 &00035880082&	2018-12-31& 1079&00088910002&	2019-08-18& 1676   \\
00035880021&	2018-08-16& 499	 &00035880083&	2019-01-01& 534	&00088910003&	2019-08-21& 1960   \\
00035880022&	2018-08-19& 349	 &00035880084&	2019-01-05& 566	&00045604010&	2019-08-29& 559    \\
00035880023&	2018-08-20& 979	 &00035880085&	2019-01-11& 862	&00045604011&	2019-09-05& 952    \\
00035880025&	2018-08-21& 748	 &00035880086&	2019-01-14& 987	&00045604012&	2019-09-12& 822    \\
00035880026&	2018-08-23& 969	 &00035880087&	2019-01-17& 1007&00045604013&	2019-09-19& 839    \\
00035880027&	2018-08-24& 1069 &00035880088&	2019-01-20& 993	&00045604014&	2019-09-26& 912    \\
00035880028&	2018-08-25& 944	 &00035880089&	2019-01-23& 1064&00045604015&	2019-10-03& 511    \\
00035880029& 	2018-08-26& 1044 &00035880090&	2019-01-26& 1020&00045604016&	2019-10-10& 920    \\
00035880030&	2018-08-27& 969	 &00035880091&	2019-01-29& 1020&00045604017&	2019-10-17& 997    \\
00035880031&	2018-08-28& 1189 &00035880092&	2019-02-02& 396	&00045604020&	2019-11-09& 1313   \\
00035880032&	2018-08-29& 1034 &00035880094&	2019-02-05& 676	&00045604021&	2019-11-14& 646    \\
00035880033&	2018-08-30& 1009 &00035880095&	2019-02-07& 1004&00045604022&	2019-11-21& 738    \\
00035880034&	2018-08-31& 939	 &00035880096&	2019-02-10& 987	&00045604023&	2019-11-28& 463    \\
00035880035&	2018-09-01& 959	 &00035880097&	2019-02-14& 1090&00045604024&	2019-12-05& 890    \\
00035880036&	2018-09-02& 959	 &00035880098&	2019-02-16& 1213&00045604025&	2019-12-12& 1075   \\
00035880037&	2018-09-03& 1019 &00035880099&	2019-02-20& 225	&00045604026&	2019-12-20& 1217   \\
00035880038&	2018-09-04& 1284 &00035880100&	2019-02-23& 888	&00045604027&	2020-01-01& 668.   \\
00035880039& 	2018-09-04& 984	 &00035880101&	2019-03-01& 947	&00045604028&	2020-01-08& 1118   \\
00035880040&	2018-09-05& 984	 &00035880102&	2019-03-04& 1010&00045604031&	2020-01-29& 669    \\
00035880041&	2018-09-06& 1059 &00035880103&	2019-03-07& 1007&00045604032&	2020-02-05& 699    \\
00035880042&	2018-09-13& 909	 &00035880104&	2019-03-10& 1183&00045604033&	2020-02-12& 870    \\
00035880043&	2018-09-17& 55	 &00035880105&	2019-03-13& 997	&00045604035&	2020-02-24& 1061   \\
00035880044&	2018-09-18& 1329 &00035880106&	2019-03-16& 862	&00045604036&	2020-02-26& 648    \\
00035880045&	2018-09-19& 964	 &00035880107&	2019-03-19& 870	&00045604037&	2020-03-04& 950    \\
00035880046&	2018-09-21& 1000 &00035880108&	2019-03-30& 724	&00045604038&	2020-03-11& 458    \\
00035880047&	2018-09-23& 1013 &00035880109&	2019-04-02& 822	&00045604040&	2020-03-25& 496    \\
00035880048&	2018-09-25& 607	 &00035880110&	2019-04-09& 679	&00045604039&	2020-03-25& 528    \\
00035880049&	2018-09-27& 966	 &00035880111&	2019-04-16& 1012&00045604041&	2020-04-08& 922    \\
00035880050&	2018-09-29& 965	 &00035880112&	2019-04-23& 1002&00045604042&	2020-04-22& 848    \\
00035880051&	2018-10-01& 992	 &00035880113&	2019-04-30& 371	&00045604043&	2020-05-06& 781    \\
00035880053&	2018-10-04& 2139 &00035880114&	2019-05-07& 684	&00045604044&	2020-05-20& 531    \\
00035880054&	2018-10-08& 1064 &00035880115&	2019-05-14& 1063&00045604045&	2020-06-03& 1060   \\
00035880055&	2018-10-09& 989	 &00035880116&	2019-05-21& 1055&00045604046&	2020-06-17& 632    \\
00035880056&	2018-10-11& 986	 &00035880117&	2019-05-28& 1012&00045604047&	2020-07-01& 1769   \\
00035880057&	2018-10-13& 967	 &00035880119&	2019-05-29& 1015&00045604048&	2020-07-15& 170    \\
00035880058&	2018-10-15& 1125 &00035880120&	2019-05-31& 699	&00045604049&	2020-07-20& 788    \\
00035880059&	2018-10-17& 1269 &00035880121&	2019-06-01& 999	&           &             &         \\
\hline
\end{tabular}
\end{table*}


\bsp	
\label{lastpage}
\end{document}